\documentclass[preprint,showpacs,tightenlines,preprintnumbers,prc,amsmath,amssymb]{revtex4}
\usepackage{graphicx}% Include figure files
\usepackage{epsfig}
\usepackage{dcolumn}% Align table columns on decimal point
\usepackage{bm}% bold math

\def\he3{$^3$He}
\def\al{\alpha}
\def\be{\beta}
\def\ga{\gamma}

\def\De{\Delta}

\def\la{\lambda}

\def\si{\sigma}

\def\w{\omega}

\def\bra{\langle}
\def\ket{\rangle}

\def\cpt{$\chi$PT }

\def\la{\lambda}

\def\calO{\mathcal{O}}
\newcommand{\lsim}{\, \, \raisebox{-0.8ex}{$\stackrel{\textstyle <}{\sim}$ }}

\newcommand{\mpi}{m _\pi}
\newcommand{\sss}{\scriptscriptstyle }

\newcommand{\vsigone}{{\vec\sigma^{\sss 1}}}
\newcommand{\vsigtwo}{{\vec\sigma^{\sss 2}}}
\newcommand{\veps}{{\hat\epsilon}}
\newcommand{\vepsprime}{{\hat\epsilon\, '}}
\newcommand{\vkay}{{\vec k}}
\newcommand{\vkayprime}{{{\vec k}\, '}}
\newcommand{\vpee}{{\vec p}}
\newcommand{\vpeeprime}{{{\vec p}\, '}}
\newcommand{\vsigma}{{\vec\sigma}}

\newcommand{\barf}{\Upsilon}

\def\vk{\hat{k}}
\def\vkp{{\hat{k}\, '}}

\def\3d{3-D}

\begin{document}

\title{Analyzing the Effects of Neutron Polarizabilities in Elastic Compton Scattering off \he3}% Force line breaks with \\

\author{Deepshikha Shukla$^1$}\email{choudhur@gwu.edu}
\author{Andreas Nogga$^2$}\email{a.nogga@fz-juelich.de}
\author{Daniel R. Phillips$^3$}
 \email{phillips@phy.ohiou.edu}
\affiliation{$^1$Department of Physics, George Washington
University, Washington DC 20052,  \\
$^2$Institut f$\ddot{u}$r Kernphysik and J$\ddot{u}$lich Center for
Hadron Physics, Forschungszentrum
J$\ddot{u}$lich, J$\ddot{u}$lich, Germany, \\
$^3$Department of
Physics and Astronomy, Ohio University, Athens, OH 45701.}

\date{\today}

\begin{abstract}
Motivated by the fact that a polarized \he3~ nucleus behaves as an
`effective' neutron target, we examine manifestations of neutron
electromagnetic polarizabilities in elastic Compton scattering from
the Helium-3 nucleus. We calculate both unpolarized and
double-polarization observables using chiral perturbation theory to
next-to-leading order (${\mathcal O}(e^2 Q)$) at energies, $\w \lsim
\mpi$, where $\mpi$ is the pion mass. Our results show that the
unpolarized differential cross section can be used to measure
neutron electric and magnetic polarizabilities, while two
double-polarization observables are sensitive to different linear
combinations of the four neutron spin polarizabilities.
\end{abstract}

\maketitle

\section{Introduction}
\label{sec:intro}

The response of an object -- that has sub-structure -- to a
quasi-static electromagnetic (EM) field is characterized in terms of
quantities called polarizabilities. For example, when an object,
having charged constituents, is placed in an electrostatic field the
centers of positive and negative charge separate resulting in an
induced electric dipole moment, $\vec p=\alpha \vec E$, where $\vec
E$ is the external electrostatic field. The strength of this
response, $\alpha$, is defined as the ``electric polarizability".
Similarly, we can imagine a magnetic dipole moment $\vec {\mu}=
\beta \vec H$ being induced in the presence of an external magnetic
field, $\vec H$, and this magnetic response is quantified in terms
of the magnetic polarizability, $\beta$. If an object has intrinsic
spin, then in the presence of an EM field the orientation of the
spin may be altered. This is the spin-dipole response; in fact, spin
responses are quantified  in terms of four spin polarizabilities.
Here, in this work we shall focus on the electric, magnetic and the
spin polarizabilities of the neutron. Investigating the neutron
polarizabilities would contribute immensely to our understanding of
the neutron structure and also proton-neutron as isospin-doublet. We
investigate neutron polarizabilities by calculating elastic Compton
scattering off \he3~ because there are no free neutron targets and
\he3~ behaves as an `effective' neutron (as we shall see later).

For the following discussion we consider both the proton and the
neutron, which we commonly call nucleon. Using the concept of an
effective theory, the most important contributions for the
interaction of a nucleon with an EM field can be identified. The
interaction Hamiltonian can then be formulated in terms of the EM
fields and nucleon EM moments. The symmetries of the EM interaction
ensure that only those terms need to be considered that obey the
following rules~\cite{barry}:
\begin{enumerate}
\item Only terms quadratic in $\vec{A}$ (the EM vector potential) are
allowed.
\item The Hamiltonian is gauge invariant,
\item a rotational scalar, and,
\item P (parity), T (time-reversal) even.
\end{enumerate}
Defining $\w$ as the photon energy, the leading-order response in
$\w$ for a nucleon in an EM field is given by--
\begin{equation}
H^{(0)}_\mathrm{eff}=\frac{(\vec p - e\vec A)^2}{2M} + e \phi,
\end{equation}
where $\vec A$ and $\phi$ are the EM vector and scalar potentials.
At the next order the following terms appear in the Hamiltonian (see
e.g.~\cite{barry, Ba98}):
\begin{equation}
H^{(2)}_\mathrm{eff}=-\frac{1}{2}\,4\pi\left[{\alpha}_{E1}\,\vec{E}^2+
                                       {\beta}_{M1}\,\vec{B}^2\right].
\label{eq:Heff}
\end{equation}
Here, ${\alpha}_{E1}$ and ${\beta}_{M1}$ are the electric and
magnetic dipole polarizabilities and are the most prominent of the
nucleon polarizabilities. This Hamiltonian (Eq.~(\ref{eq:Heff})),
however, includes only terms of second order in $\omega$ and can be
extended, as long as the aforementioned conditions are respected.
Apart from $\alpha_{E1}$ and $\beta_{M1}$ there exist other
polarizabilities, such as the spin-dependent polarizabilities.
Therefore, we can modify and rewrite Eq.~(\ref{eq:Heff}), extended
to the spin dipole polarizabilities~\cite{barry, Ba98}:
\begin{eqnarray}
H^{(3)}_\mathrm{eff}&=&-2\pi\,\left[\alpha_{E1}\,\vec{E}^2+
\beta_{M1}\,\vec{B}^2+
\gamma_{E1E1}\,\vec{\sigma}\cdot\vec{E}\times\dot{\vec{E}} +
 \gamma_{M1M1}\,\vec{\sigma}\cdot\vec{B}\times\dot{\vec{B}}\right. \nonumber\\
&& - \left. 2\,\gamma_{M1E2}\,\sigma_i\,E_{ij}\,B_j  +
2\,\gamma_{E1M2}\,\sigma_i\,B_{ij}\,E_j\right], \label{eq:Heff2}
\end{eqnarray}
with $\vec{\sigma}$ denoting the intrinsic nucleon spin. The various
subscripts in $\al$, $\be$ and $\ga$ denote different multipoles of
the nucleon's response to the external EM field and
\begin{eqnarray}
E_{ij}&=& \frac{1}{2}(\nabla_i E_j + \nabla_j E_i) \\
B_{ij}&=& \frac{1}{2}(\nabla_i H_j + \nabla_j H_i).
\end{eqnarray}
In this work, we use $\ga_1$, $\ga_2$, $\ga_3$ and
$\ga_4$~\cite{ragusa} to represent the four spin polarizabilities
and these are related to the $\ga$'s in Eq.~(\ref{eq:Heff2}) as --
\begin{eqnarray}
\gamma_{E1E1}&=& -\ga_1 - \ga_3, \nonumber \\
\gamma_{M1M1}&=& \ga_4, \nonumber \\
\gamma_{M1E2}&=& \ga_2 + \ga_4, \nonumber \\
\gamma_{E1M2}&=& \ga_3.
\end{eqnarray}

An electromagnetic probe by nature, Compton scattering captures
information about the response of the charge and current
distributions inside a nucleon, and hence the polarizabilities, to a
quasi-static electromagnetic field. To lowest order in photon energy
(${\mathcal O}(\w^0)$), the spin-averaged amplitude for Compton
scattering on the nucleon is given by the Thomson term
\begin{equation}
{\rm Amp}=-\frac{{\mathcal Z}^2
e^2}{M}\hat{\epsilon}\cdot\hat{\epsilon}', \label{eq:1}
\end{equation}
where ${\mathcal Z}e,M$ represent the nucleon charge and mass
respectively and $\hat{\epsilon},\hat{\epsilon}'$ specify the
polarization vectors of the initial and final photons respectively.
At the next order in photon energy, contributions arise from
electric and magnetic polarizabilities---$\bar{\alpha}_E$ and
$\bar{\beta}_M$---which measure the response of the nucleon to the
application of quasi-static electric and magnetic fields. The
spin-averaged amplitude is expressed in the laboratory frame as:
\begin{equation}
{\rm Amp}=\veps\cdot\vepsprime\left(-\frac{{\mathcal Z}^2 e^2}{M}
+\omega\omega' \; 4\pi\bar{\alpha}_E \right)
+\veps\times\vk\cdot\vepsprime \times\vkp\; \omega\omega'
4\pi\bar{\beta}_M+\; {\mathcal O}(\omega^4) \; . \label{eq:2}
\end{equation}
Here, $k_\mu=(\omega,\vec{k})$, ${k'}_\mu=(\omega',\vec{k}')$
specify the four-momenta of the initial and final photons
respectively and $\vk$ and $\vkp$ are unit vectors associated with
the photon momenta. When we square the amplitude~(\ref{eq:2}), the
associated differential scattering cross section on the proton is
given by
\begin{eqnarray}
{d\sigma\over d\Omega}&=&\left({e^2\over 4\pi M}\right)^2
\left({\omega'\over \omega}\right)^2\left[{1\over 2}
(1+\cos^2\theta)\right.\nonumber\\
&-&\left.{4\pi M\omega\omega'\over e^2}\left({1\over 2}
(\bar{\alpha}_E+\bar{\beta}_M)(1+\cos\theta)^2+{1\over
2}(\bar{\alpha}_E-\bar{\beta}_M)(1-\cos\theta)^2\right)+\ldots\right].
\nonumber\\
\label{labdcs} \quad
\end{eqnarray}
From Eq.~(\ref{labdcs}) it is evident that the differential
cross-section is sensitive to $(\bar{\alpha}_E+\bar{\beta}_M)$ at
forward angles and to $(\bar{\alpha}_E-\bar{\beta}_M)$ at backward
angles.

It is essential to mention here that for the purpose of this work,
$\bar{\alpha}_E$ and $\bar{\beta}_M$ are the so-called Compton
polarizabilities and measure the ``true deformation" effect on the
nucleon. In other words, Compton polarizabilities measure the
distortion effects of charge and current distributions inside a
nucleon subjected to an EM field. In the non-relativistic limit,
these reduce to the static electromagnetic polarizabilities
described in Eq.~(\ref{eq:Heff}). For example, the static electric
polarizability, $\alpha_{E1}=\bar{\alpha}_E + {\kappa^2 \over M}$
\cite{Ba97}, where $\kappa$ is the anomalous magnetic moment of the
nucleon and $M$ is its mass. From here on we shall drop the
subscripts $E$ and $M$ as well as the bars over $\al$ and $\be$ from
the Compton polarizabilities. Therefore all polarizabilities
discussed from now on are the Compton polarizabilities. At the next
order in $\w$ beyond that considered in Eq.~(\ref{eq:2}), one can
access the spin polarizabilities ($\ga$'s) of the nucleon.

The last fifteen years or so has been witness to a concerted effort
on the part of both theorists and experimentalists to comprehend the
structure of nucleons as manifested in Compton scattering. Through
various experiments on the proton, its electric and magnetic
polarizabilities have been effectively pinned down. There has been
an avalanche of experimental data on unpolarized Compton scattering
on the proton at photon energies below 200 MeV~\cite{Fe91, Ze92,
Ha93, Ma95, Ol01}. The current Particle Data Group (PDG) values for
the proton are:
\begin{eqnarray}
\alpha_p&=&(12.0 \pm 0.6) \times 10^{-4} \, {\rm fm}^3, \nonumber \\
\beta_p&=& (1.9 \pm 0.5) \times 10^{-4} \, {\rm fm}^3.
\label{eq:pexp}
\end{eqnarray}
Most of the extractions of the proton electromagnetic
polarizabilities were performed via a dispersion relation approach
by a multipole analysis of the photoabsorption
amplitudes~\cite{Ar96, Lv97}. Calculations have also been done up to
$\calO (e^2 Q^2)$ in \cpt~\cite{McG01, Be02, Be04} and these
calculations give a very impressive description of the data for
$\w$, $\sqrt{|t|}<$200~MeV.

Note that it is necessary to measure only one of $\al$ and $\be$. In
practice, one can then use the dispersion sum rule (Baldin sum rule)
derived from the optical theorem~\cite{bkmrev}
\begin{equation}
\al + \be = \frac{1}{2 \pi^2} \int_{\w_{th}}^{\infty}
\frac{\si_p^{tot} (\w)}{\w^2}\, \mathrm{d}\w,
\label{baldin}
\end{equation}
that relates the sum of the polarizabilities to the total
photoabsorption cross-section in order to extract the other. In
Eq.~(\ref{baldin}), $\si_p^{tot}(\w)$ is the total photoabsorption
cross-section and $\w_{th}$ is the pion-production threshold. For
the proton~\cite{Ol01},
\begin{equation}
\al_p + \be_p  = (13.8 \pm 0.4) \times 10^{-4} \, {\rm fm}^3.
\label{eq:baldinp}
\end{equation}

There are several evaluations of the dispersion relation in
Eq.~(\ref{baldin}) for the neutron~\cite{nbaldin, Ba98, Lv00}.
However, there is discrepancy between the numbers extracted.
%The
%numbers for the sum-rule do not agreeIn Ref.~\cite{Ba98} the sum
%rule for the neutron was extracted from the deuteron photoabsorption
%cross-section by subtracting the proton contribution.
%They obtained,
%\begin{equation}
%\al_n + \be_n = (14.4 \pm 0.66) \times 10^{-4} \, {\rm fm}^3.
%\label{eq:baldinn1}
%\end{equation}
A recent extraction of the sum rule by Levchuk and L'vov~\cite{Lv00}
using relevant neutron photo-production multipoles gives
\begin{equation}
\al_n + \be_n = (15.2 \pm 0.5) \times 10^{-4} \, {\rm fm}^3.
\label{eq:baldinn2}
\end{equation}
Because of the discrepancy in the extraction of the neutron sum
rule, as far as the neutron polarizabilities go, we still strive to
extract precise numbers for the polarizabilities. Since neutrons are
very short-lived, we lack free neutron targets and this poses a
serious handicap in the study of neutron structure. The drawback
caused by not having free neutron targets encouraged the community
to look at other avenues to extract information about the neutron.

There were attempts to measure neutron polarizabilities via
scattering neutrons on lead to access the Coulomb field of the
target and examining the cross-section as a function of energy.
Currently there is much controversy over what this technique gives
for $\al_n$. Two experiments using this same technique obtained
different results for $\al_n$ (Refs.~\cite{Sc91, Ko95}).
\begin{eqnarray}
\alpha_n&=&(12.6 \pm 1.5 \pm 2.0) \times 10^{-4} \, {\rm fm}^3;
\label{eq:nexp1} \\
\alpha_n&=& (0.6 \pm 5.0) \times 10^{-4} \, {\rm fm}^3.
\label{eq:nexp2}
\end{eqnarray}
Enik et al~\cite{En97} revisited Ref.~\cite{Sc91}
(Eq.~(\ref{eq:nexp1})) and recommended a value of $\al_n$ between $7
\times 10^{-4}$ fm$^3$ and $19 \times 10^{-4}$ fm$^3$.

In addition, quasi-free Compton scattering from the deuteron was
measured at SAL~\cite{Ko00a} for incident photon energies of
$E_{\ga} = (236 - 260)$ MeV and one-sigma constraints on the
polarizabilities were reported to be:
\begin{eqnarray}
\alpha_n&=&(7.6 - 14.0) \times 10^{-4} \, {\rm fm}^3,
\nonumber \\
\beta_n&=& (1.2 - 7.6) \times 10^{-4} \, {\rm fm}^3.
\label{eq:nexp3}
\end{eqnarray}
Another quasi-free experiment was performed at Mainz~\cite{Ko03} and
the difference in the polarizabilities was obtained:
\begin{equation}
\al_n-\be_n=(9.8 \pm 3.6 {\rm (stat)} \pm 2.2 {\rm (model)}
{}^{+2.1}_{-1.1} {\rm (sys)}) \times 10^{-4} \, {\rm fm}^3.
\label{mainzn}
\end{equation}

Until now, however, there have been only limited efforts to measure
nucleon spin polarizabilities. The only ones that have been
extracted from the experimental data are the forward and backward
spin polarizabilities. The backward spin polarizability is defined
as:
\begin{equation}
\ga_{\pi} = \ga_1+\ga_2+2\ga_4.
\end{equation}

The neutron backward spin polarizability was determined to be
\begin{equation}
\ga_{\pi n} = (58.6 \pm 4.0) \times  10^{-4} \, {\rm fm}^4,
\label{eq:gpn}
\end{equation}
from quasi-free Compton scattering on the deuteron~\cite{Ko03}. This
experiment used the Mainz 48 cm $\varnothing \times$ 64 cm NaI
detector and the G\"{o}ttingen recoil detector SENECA in
coincidence. For comparison, an experiment~\cite{Ca02} using the
same detector set-up extracted a proton backward polarizability
value ranging from (-36.5 to -39.1)$\times 10^{-4}$ fm$^4$. These
values are consistent with earlier Mainz measurements~\cite{Ol01,
Ga01, Wo01}.

The theoretical prediction for $\ga_{\pi p}$ from \cpt is $-36.7
\times 10^{-4}$ fm$^4$~\cite{He98, Pa03} and these calculations were
done up to the next-to-leading order with the $\Delta$-isobar as an
explicit degree of freedom. The prediction \cpt for $\ga_{\pi n}$ is
$57.4 \times 10^{-4}$ fm$^4$~\cite{Pa03}. Both predictions are in
agreement with the experimental numbers to within their quoted
uncertainties.

The forward spin polarizability, $\ga_0$, is related to
energy-weighted integrals of the difference in the
helicity-dependent photoreaction cross-sections ($\si_{1/2} -
\si_{3/2}$). Using the optical theorem one can derive the following
sum rule for the forward spin polarizability~\cite{bkmrev,ggt}:
\begin{equation}
\ga_0 = \ga_1 - (\ga_2 + 2\ga_4) =  \frac{1}{4 \pi^2}
\int_{\w_{th}}^{\infty} \frac{\si_{1/2}-\si_{3/2}}{\w^3}\,
\mathrm{d}\w, \label{eq:g0}
\end{equation}
where $\w_{th}$ is the pion-production threshold. The following
results on $\ga_0$ were estimated using the VPI-FA93 multipole
analysis~\cite{Sa94} to calculate the integral on the RHS of
Eq.~(\ref{eq:g0}):
\begin{eqnarray}
\ga_{0p} &\simeq& -1.34 \times 10^{-4} \, {\rm fm}^4,
\label{eq:g0p} \\
\ga_{0n} &\simeq& -0.38 \times 10^{-4} \, {\rm fm}^4. \label{eq:g0n}
\end{eqnarray}
The \cpt prediction for $\ga_{0p}$ is $-2.1 \times 10^{-4}$
fm$^4$~\cite{Pa03} and for $\ga_{0n}$ is consistent with
0~\cite{Pa03}.

Let us now summarize the ``state of the nucleon polarizabilities".
The quantities $\al_p$ and $\be_p$ are well known, $\al_n$ and
$\be_n$ are also known. However, looking at the numbers in
Eqs.~(\ref{eq:baldinn2})--(\ref{mainzn}), it is evident that the
extractions of $\al_n$ and $\be_n$ have large error bars.
Strikingly, there is no consensus on the numbers for $\ga_{1p}
\ldots \ga_{4p}$ or on $\ga_{1n} \ldots \ga_{4n}$. Of these,
$\ga_{1p} \ldots \ga_{4p}$ are not known due to a lack of
experimental data. But it is clear that the extraction of $\al_n$,
$\be_n$ and $\ga_{1n} \ldots \ga_{4n}$ require efforts from both the
theoretical and the experimental community.

As there are no free neutron targets, the theoretical community has
directed focus on elastic Compton scattering from light nuclei to
access the neutron polarizabilities. The main advantage of such a
reaction process compared to the more complex n-Pb scattering is
that the theoretical analysis of the few-body reactions is much
better controlled. The lightest nucleus is the deuteron and elastic
$\gamma d$ scattering has been studied with the intent of extracting
information about the neutron polarizabilities from unpolarized and
polarization observables. It should be noted, however, that
processes with $A>1$ pose a different kind of challenge because they
require an understanding of the inter-nucleon interaction. Deuteron
structure is governed by the NN interaction and hence, when
analyzing $\ga$d data, one has to include the effects of photons
coupling to mesons being exchanged between nucleons (`two-body
currents') over and above the single-nucleon $\gamma$N amplitude.
There exist calculations for $\ga$d scattering in the framework of
conventional potential models~\cite{We83, Wi95a, Lv95, Lv98, Lv00,
Ka99}. All of these calculations use a realistic NN interaction and
usually give a good description of the data. However, the
predictions of these calculations depend somewhat on the NN model
chosen. \cpt on the other hand allows a model independent framework
to systematically build the theory for $\ga$d scattering. The
amplitude for coherent Compton scattering on the deuteron has been
calculated to $\calO (e^2 Q^2)$ by Beane et. al.~\cite{Be02, Be04}.
At this order there are four new parameters in the theory (two of
which can be fit to the $\ga p$ data) and since the deuteron is an
isoscalar, the nucleon isoscalar polarizabilities were extracted
from the above experimental data by fitting the other two
parameters. The isoscalar polarizabilities are linear combinations
of the proton and neutron polarizabilities, $\al_N \equiv {(\al_p
+\al_n) \over 2}$ and $\be_N \equiv {(\be_p +\be_n) \over 2}$. Beane
$et \, al.$ obtained--
\begin{eqnarray}
\alpha_N&=& (8.9 \pm 1.5)^{+4.7}_{-0.9} \times 10^{-4} \, {\rm
fm}^3,
\nonumber \\
\beta_N&=& (2.2 \pm 1.5)^{+1.2}_{-0.9} \times 10^{-4} \, {\rm fm}^3.
\label{eq:oq4pol}
\end{eqnarray}
Subsequently, next-to-leading order calculations for deuteron
Compton scattering have been performed that include the $\De$ as an
explicit degree of freedom and also incorporate the low-energy NN
rescattering contributions in the intermediate state~\cite{Hi05a,
Hi05b, Hi05}. Thus, sophisticated calculations on $\ga$d scattering
already exist and with more experimental data soon to come from
MAXLab in Sweden there is hope that $\al_n$ and $\be_n$ can be
extracted with good precision from that experiment.

Recent experimental advances (for example technology to polarize
targets) have made it possible to venture into the area of
polarization observables. Hildebrandt $et \, al.$~\cite{Hi04b, Hi05}
have also studied polarization observables for Compton scattering on
the nucleon (both proton and neutron) with a focus on the spin
polarizabilities. Their calculation involves free nucleons and may
be beneficial for understanding processes that involve quasi-free
kinematics like $\ga d \rightarrow \ga np$. Choudhury and
Phillips~\cite{Ch05, mythesis} focused on $\vec{\ga} d \rightarrow
\ga d$ and $\vec{\ga} \vec{d} \rightarrow \ga d$ to analyze the
effects of the electric, magnetic and the spin polarizabilities of
the neutron. They reported that one of the double-polarization
observables was sensitive to a linear combination of the neutron
spin polarizabilities and the deuteron Compton scattering
observables alone were not enough to extract the neutron
polarizabilities. Calculations for polarization observables for
deuteron Compton scattering with explicit $\Delta$-isobar and
low-energy resummation have also been completed and the results are
forthcoming~\cite{Sh08}.

These studies~\cite{We83, Wi95a, Lv95, Lv98, Lv00, Ka99, Be99, Be02,
Be04,Hi05a, Hi05b, Hi04b, Hi05, Ch05, mythesis, Sh08} make it
evident that not all of the neutron polarizabilities can be
extracted from deuteron Compton scattering. This means that several
different observables or combinations of observables from different
reactions are necessary to effectively pin down the neutron
polarizabilities. To this end, physicists should focus on other
alternatives, for instance, quasi-free kinematics in deuteron
Compton scattering, or other nuclear targets to extract information
about the neutron. Such parallel calculations to study the effect of
the same neutron polarizabilities combined with experiments can
build confidence in the extraction of the polarizabilities. In
principle, one should extract the same numbers, no matter what the
process or the target is. Thus, these parallel studies would
reassure us about our understanding of ``nuclear" effects.

\begin{figure}[htbp]
\epsfig{figure=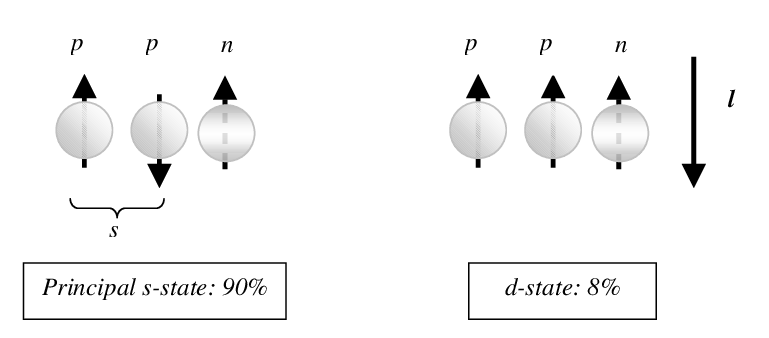, height=2in} \caption [The polarized $\,
^3He $ nucleus.]{The polarized \he3~ nucleus. 90\% of the time the
nucleus is in the principal s-state configuration. \vspace{0.4in}
\label{fig:he3pol}}
\end{figure}

One alternative target is the polarized \he3~ nucleus. This nucleus
has the nice property that the two proton spins are anti-aligned for
the most dominant part of the \he3~ wavefunction, the principal
`$s$-state'. Since approximately 90\% of the wavefunction is given
by this state, the spin of \he3~ nucleus is mostly carried by the
unpaired neutron alone. Fig.~\ref{fig:he3pol} shows the
configurations for the principal `$s$-state' and for one other
possible contributions to the \he3~ wavefunction~\cite{he3pol,
No03}. This paper reports our calculations for Compton scattering on
a \he3~ nucleus.
Note that, for $\ga$\he3~ scattering we now have to understand the
interplay of the three nucleons over and above the two-nucleon
effects.

In this work, we calculate {\it elastic} Compton scattering on $\,
^3$He to $\calO(e^2 Q)$ and focus on specific observables so as to
construct road-maps to extract the neutron polarizabilities,
$\al_n$, $\be_n$ and $\ga_{1n} \ldots \ga_{4n}$. It should be
mentioned here that these are the first ~\he3 Compton scattering
calculations in any framework. These calculations include the
processes $\ga ^3$He$ \rightarrow \ga ^3$He and $\vec{\ga}
\vec{^3He} \rightarrow \ga ^3$He and are necessarily exploratory.
They are expected to provide benchmark results for elastic Compton
scattering on $\, ^3$He. As such, our results will serve to attract
further explorations of these observables at $\calO(e^2 Q^2)$ and
beyond. The main conclusions of this work have already been reported
in Ref.~\cite{long, mythesis}. This paper is intended to be a
comprehensive discussion on the calculations that lead to these
conclusions.

This paper is organized in the following manner. Sec.~\ref{sec:anat}
lays out the structure of our calculation. In Sec.~\ref{sec:obs} we
describe the various observable that we focus on and then
Sec.~\ref{sec:res} we report our results. In order to make this work
comprehensive, a discussion on inherent sources of uncertainties
that may affect the final result is provided in Sec.~\ref{sec:err}.
Finally, in Sec.~\ref{sec:sum}, the calculation is summarized and
also future directions for calculations of $\ga$\he3~ scattering are
pointed out.

\section{Anatomy of the Calculation}
\label{sec:anat}

\begin{figure}[htbp]
\epsfig{figure=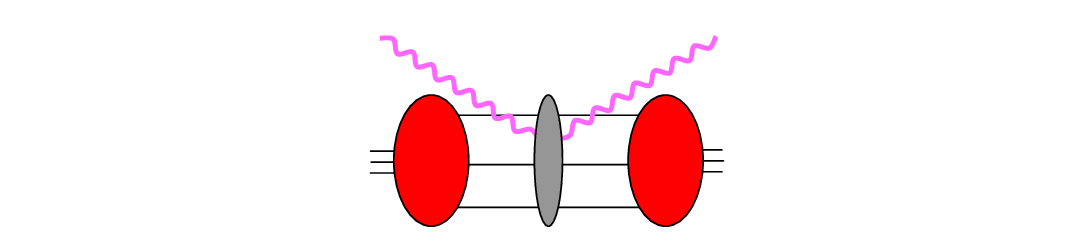, height=1 in} \caption {The anatomy of
the calculation. The irreducible amplitude is computed in HB\cpt and
sandwiched between external \he3~ wavefunctions to give the matrix
element for Compton scattering on \he3~. \vspace{0.4in}
\label{fig:ana}}
\end{figure}

\begin{figure}[htbp]
\epsfig{figure=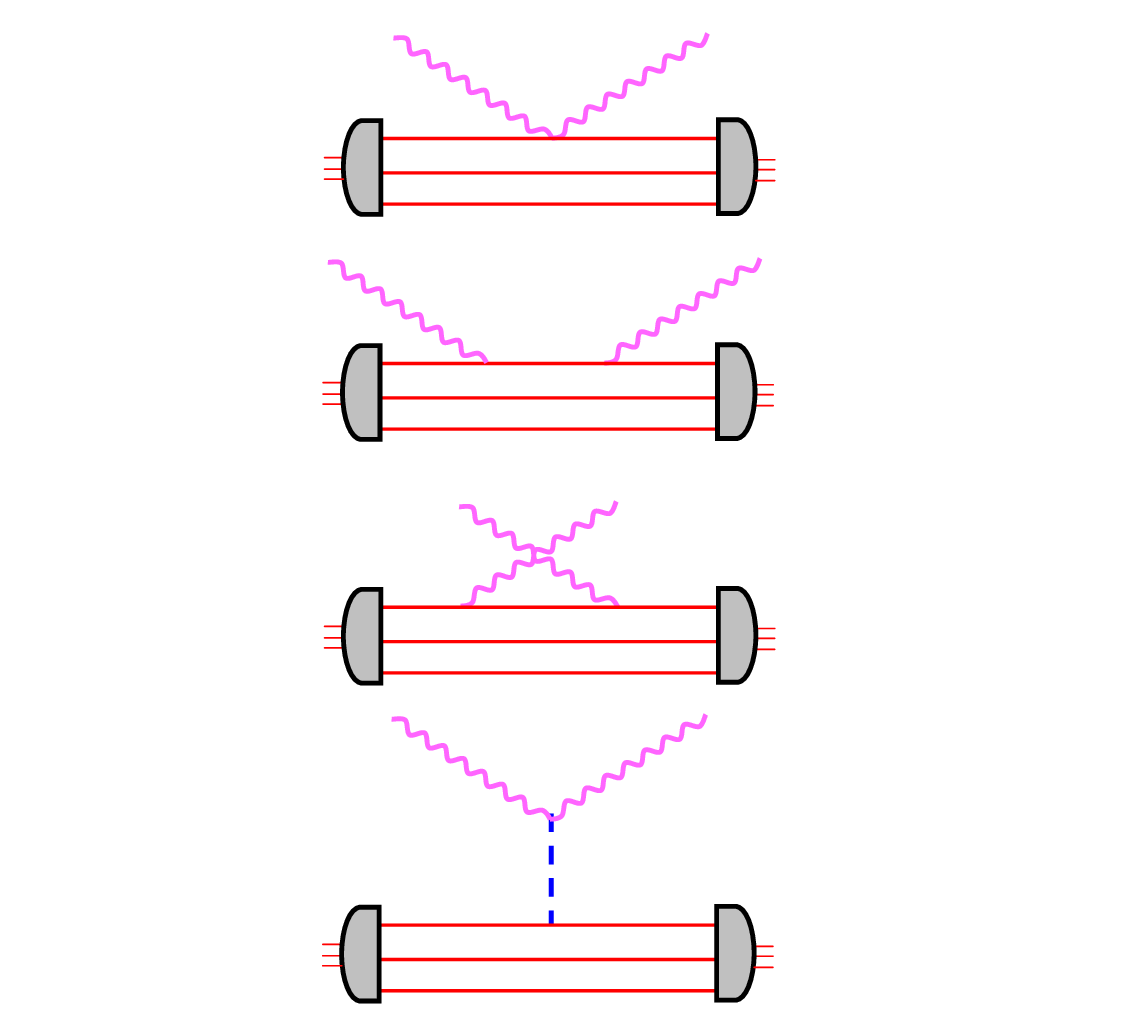, height=4in} \caption {Tree level
diagrams for the one-body amplitude. \vspace{0.3in}
\label{fig:tree}}
\end{figure}

\begin{figure}[htbp]
\epsfig{figure=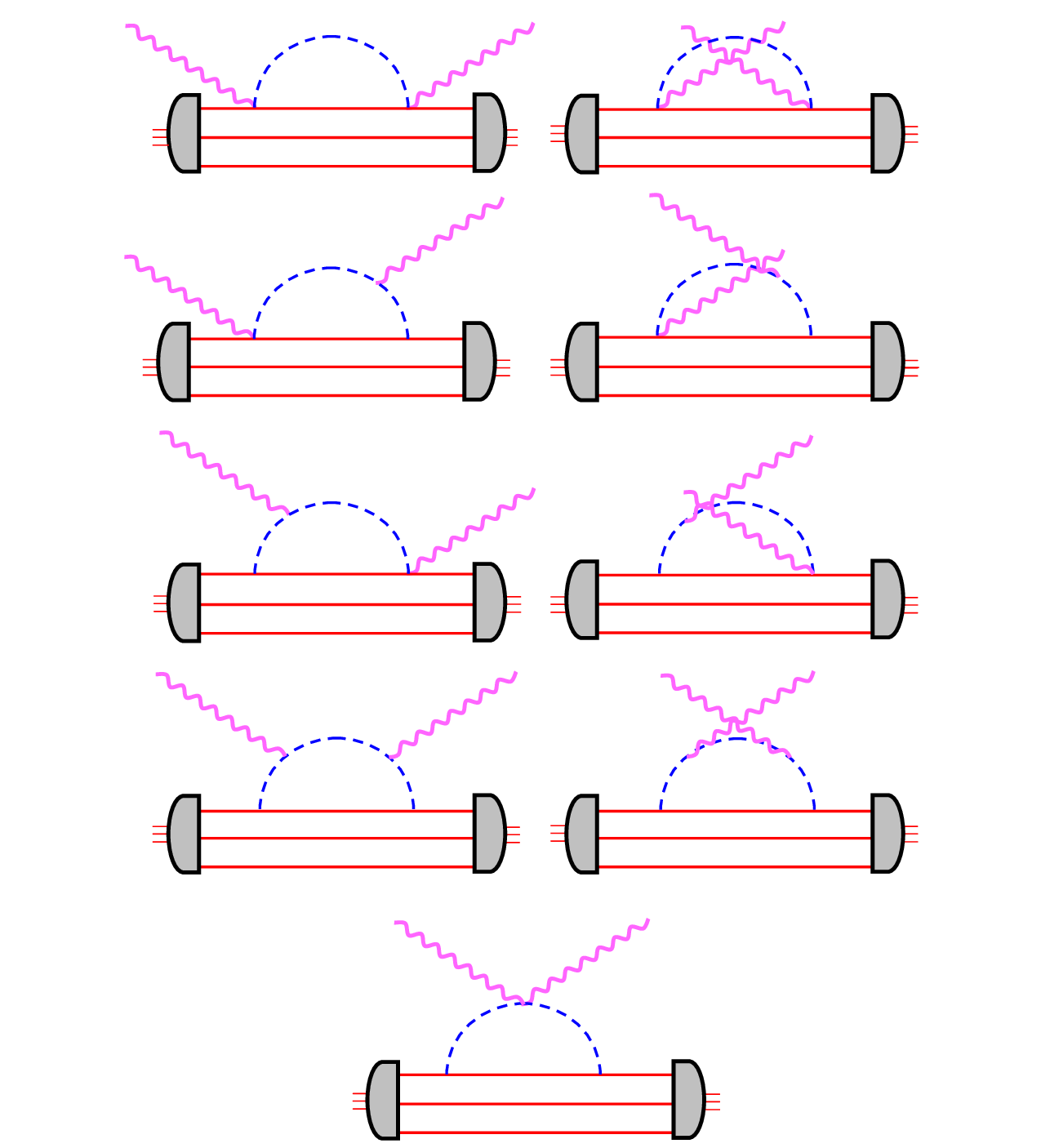, height=5in}  \caption {Contribution to
the one-body amplitude at $\calO(e^2 Q)$. These diagrams contain one
pion loop. \label{fig:IA}}
\end{figure}

\begin{figure}[htbp]
\epsfig{figure=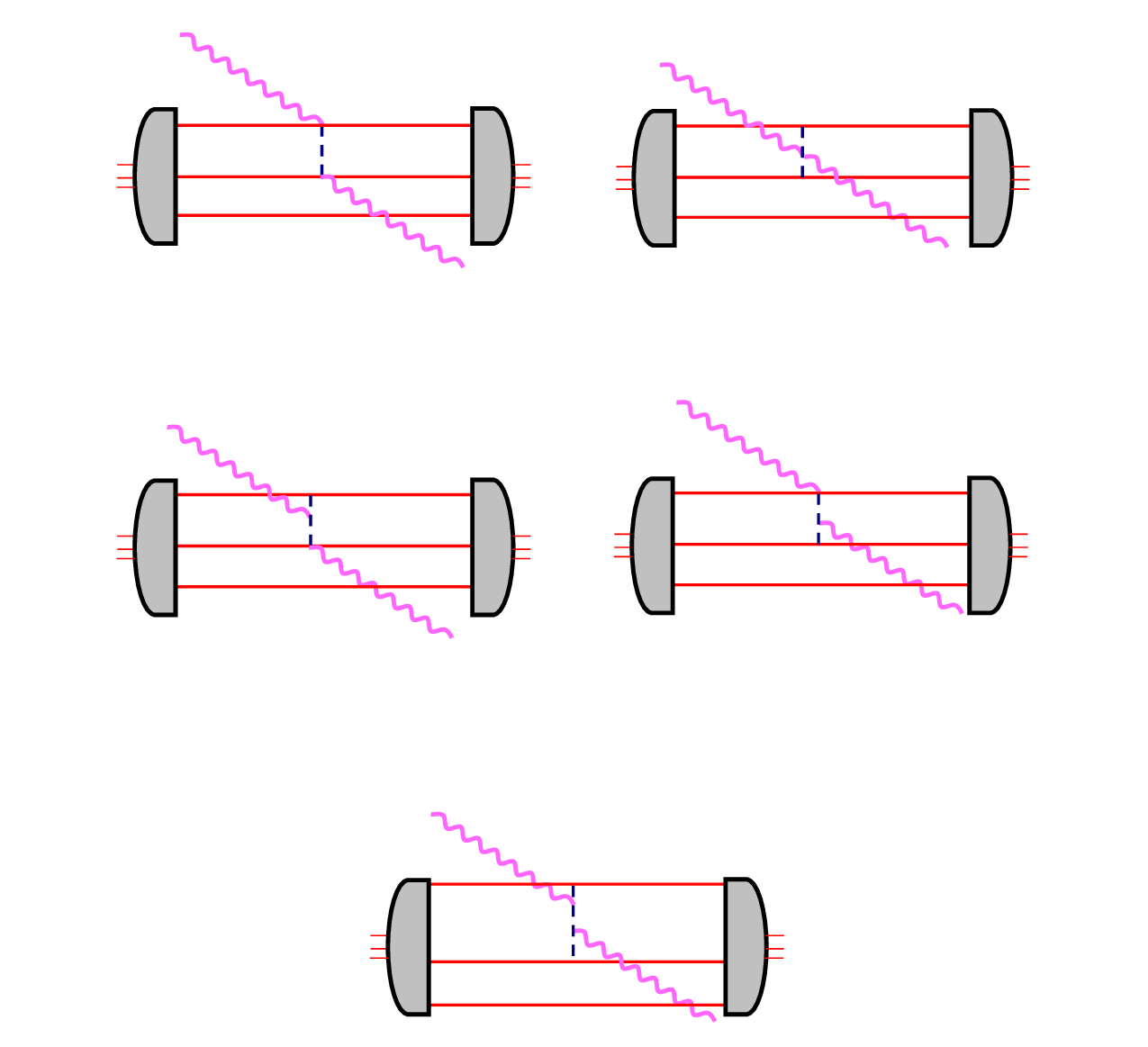, height=4in} \caption {Contributing
two-body diagrams at $\calO(e^2 Q)$. Permutations are not shown.
\vspace{0.3in} \label{fig:2B}}
\end{figure}

The irreducible amplitudes for the elastic scattering of real
photons from the NNN system are first ordered and calculated in
HB\cpt. Then these amplitudes are sandwiched between the nuclear
wavefunctions to finally obtain the scattering amplitudes (see
Fig.~\ref{fig:ana}). This amplitude can be written as--
\begin{equation}
{\mathcal M}=\bra \Psi_f|{\hat O}|\Psi_i \ket \label{eq1}
\end{equation}
Here, $|\Psi_i\ket$ or $|\Psi_f\ket$ are the \he3~ wavefunctions
that are anti-symmetrized in order to take into account that the
nucleons are identical fermions. We will employ \he3~ wavefunctions
obtained from Faddeev calculations in momentum space. We first
calculate a specific Faddeev component $|\psi \ket_3$ for \he3~
which is related to the fully anti-symmetrized wavefunction $|\Psi
\ket$~\cite{No97} by
\begin{equation}
|\Psi\ket = (1+P)|\psi\ket_3
\end{equation}
where $P=P_{31}P_{12} + P_{32}P_{12}$ is the sum of the cyclic and
anti-cyclic permutation operators and the numbered subscripts denote
the nucleon number.
In our case, since we are calculating only up to $\calO(e^2 Q)$ in
HB\cpt, the operator ${\hat O}$ consists of a one-body part
\begin{equation}
{\hat O}^{1B}={\hat O}^{1B}(1)+{\hat O}^{1B}(2)+{\hat O}^{1B}(3),
\label{eq2}
\end{equation}
and a two-body part
\begin{equation}
{\hat O}^{2B}={\hat O}^{2B}(1,2)+{\hat O}^{2B}(2,3)+{\hat
O}^{2B}(3,1). \label{eq3}
\end{equation}
The mechanisms that contribute to the two different pieces are shown
in Figs.~\ref{fig:tree}, \ref{fig:IA} and \ref{fig:2B}. The
structure of the $\ga$N and the $\ga$NN amplitudes follows
Refs.~\cite{bkmrev, Be99} and they are shown in Apps.~\ref{sec:gaN}
and \ref{sec:gaNN}. In the notation used in Eq.~(\ref{eq2}), ${\hat
O}^{1B}(a), a=1\ldots 3$ represents the one-body current where the
external photon interacts with nucleon `$a$' and in Eq.~(\ref{eq3}),
${\hat O}^{2B}(a,b)$ represents a two-body current where the
external photon interacts with the nucleon pair `$(a,b)$'. In short,
the different terms in Eqs.~(\ref{eq2}) and (\ref{eq3}) represent
different permutations of the three nucleons inside \he3~. Note
that, we do not have any three-body currents because they appear
only at ${\mathcal O}(e^2 Q^3)$ in the \cpt power-counting.

\begin{figure}[htbp]
\epsfig{figure=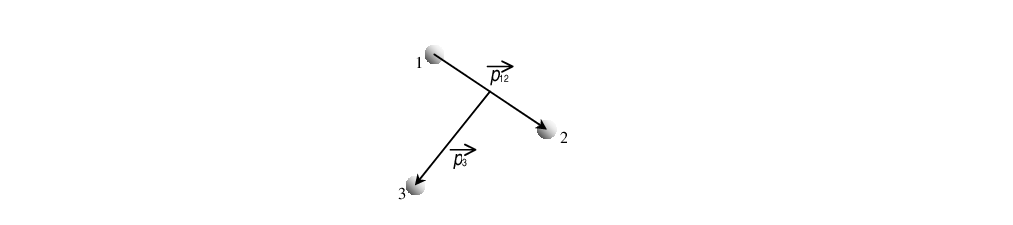, height=1.5 in} \caption {Jacobi
momenta for the three-nucleon system. \vspace{0.4in}
\label{fig:jacobi}}
\end{figure}

If we now take into account the identity of the three nucleons, we
can rewrite Eq.~(\ref{eq1}) as--
\begin{equation}
{\mathcal M} = 3\bra \Psi_f|{\hat O}^{1B}(1)+{\hat
O}^{2B}(1,2)|\Psi_i\ket \label{eq4}
\end{equation}
For our calculations we draw upon the work of Kotlyar et
al~\cite{Ko00}. They developed an approach for calculating matrix
elements of meson-exchange current operators between three-nucleon
basis states. They applied this approach in studying \he3~ photo-
and electro-disintegration. The three-nucleon basis states were
expressed in a $jj$-coupling scheme (final configuration) and a
three-nucleon bound state (initial configuration). These matrix
elements were then expressed in terms of multiple integrals in
momentum space. Since they studied photo- and
electro-disintegration, they had only one external momentum transfer
vector $\vec q$ that they could choose to align in a preferred
direction in order to simplify the calculations. Here in contrast,
since we have an incoming and outgoing photon with different
three-momenta, we do not have the same liberty and hence we needed
to extend their prescription. Also, we choose to calculate the
one-body currents too in the two-nucleon spin-isospin basis and
hence it is more convenient to express Eq.~(\ref{eq4}) as--
\begin{equation}
{\mathcal M} = 3\bra \Psi_f|\frac{1}{2} \big( {\hat O}^{1B}(1)+{\hat
O}^{1B}(2) \big)+{\hat O}^{2B}(1,2)|\Psi_i\ket = 3\bra \Psi_f|{\hat
O}(1,2)|\Psi_i\ket. \label{eq5}
\end{equation}
The advantage of this formulation is that the structure of the
calculation is similar for both the one-body and the two-body parts.
The next step is then to actually calculate the matrix elements. To
do this, we project the state-vectors and the operators on to the
basis--
\begin{equation}
|p_{12} p_3 \, \alpha\ket=|p_{12} p_3 \ket|(l_{12}s_{12})j_{12}(l_3
{1\over 2})j_3 (j_3 j_{12})JM_J\ket|(t_{12} {1\over 2})T M_T\ket_3 =
|p_{12} p_3 \ket|\alpha_J\ket|\alpha_T\ket_3. \label{eq6}
\end{equation}
Here, $p_{12}$ and $p_3$ denote the magnitude of the Jacobi momenta
of the pair ``(1,2)" and the spectator nucleon ``3" and this choice
ensures that the set of basis states are complete and
orthonormalized. The subscript `3' represents the choice of the
Jacobi momenta (Fig.~\ref{fig:jacobi}) defined as--
\begin{eqnarray}
\vec p_{12} &=& {1\over 2} (\vec k_2 - \vec k_1) \nonumber \\
\vec p_3 &=& {1\over 3}(2\vec k_3 - \vec k_2 - \vec k_1).
\label{eq7}
\end{eqnarray}
$i.e.$ nucleon `3' serves as the spectator.
The total angular momentum of the nucleus $J$ is a result of
coupling between $j_{12}$ and $j_3$, the total angular momenta of
the `(1,2)' subsystem and the spectator nucleon `3' respectively.
The orbital angular momentum $l_{12}$ and the spin $s_{12}$ of the
two-body subsystem `(1,2)' combine to give $j_{12}$ and similarly,
$l_3$ and $s_3={1\over 2}$ combine to give $j_3$. $t_{12}$ is the
isospin of the two-nucleon subsystem and it combines with the
isospin of the spectator nucleon to give the total isospin $T$ and
$M_T$ is simply the projection of $T$. Since we are concerned with
the \he3~ nucleus--
\begin{eqnarray}
|T M_T\ket &=& |{1 \over 2}{1 \over 2}\ket \nonumber
\\
|J M_J\ket &=& |{1 \over 2}M_J\ket.
\end{eqnarray}
The basis states as defined above are not completely
anti-symmetrized. The antisymmetry is carried by the complete
wavefunction. However, for the two-body subsystem quantum numbers
the antisymmetry within the subsystem leads to the additional
constraint that $l_{12}+s_{12}+t_{12}=2n+1, (n=0,1,2,\ldots)$.

Before delving into the matrix element calculations, let us first
consider the kinematics of Compton scattering from a \he3~ nucleus
with momenta assigned in this way.

\subsection{Kinematics}
\label{sec:kin}

Since we are interested in two-body currents at the most, the
kinematics of our problem becomes a little simplified. In our
calculations we shall be working in the $\ga \, ^3$He c.m. frame.
\begin{figure}[htbp]
\epsfig{figure=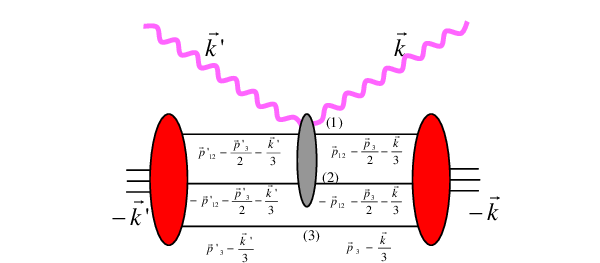, height=3in}\caption {The kinematics of
the calculation in $\gamma$\he3~ c.m. frame. The vectors $\vec{k}$
($-\vec{k}$) and $\vec{k}'$ ($-\vec{k}'$) are the three-momenta of
the incoming and outgoing photon (\he3) in the $\gamma$\he3~ c.m.
frame. \label{fig:kin}}
\end{figure}
As we can see in Fig.~\ref{fig:kin}, nucleon ``3" serves as a mere
spectator to the $\ga$NN $\rightarrow \ga$NN scattering process and
this means that in our calculations we can treat the ``(1,2)" and
``3" momentum spaces separately. Moreover, since the third nucleon
does not partake in the interaction, we have a three-momentum
conserving delta-function from--
\begin{eqnarray}
\vec p_3\,' - {\vec k' \over 3} &=& \vec p_3 - {\vec k \over 3}
\nonumber \\
\Rightarrow \vec p_3\,' &=& \vec p_3 + {(\vec k' - \vec k) \over 3}
\nonumber \\
&=& \vec p_3 + {\vec q \over 3}, \label{vecp3}
\end{eqnarray}
where, $\vec q=(\vec k' - \vec k)$. Hence, in principle, any
integral in $\vec p_3\,'$ space can be eliminated by using this
simplification. However, as is evident from Fig.~\ref{fig:kin},
there are still three unknown three-momenta and hence we have to
perform a nine-dimensional integral to calculate the matrix
elements. We shall see later that the matrix-elements pertaining to
the one-body currents are further reduced to six-dimensional
integrals because of an additional three-momentum conserving
delta-function.

\begin{figure}
\epsfig{figure=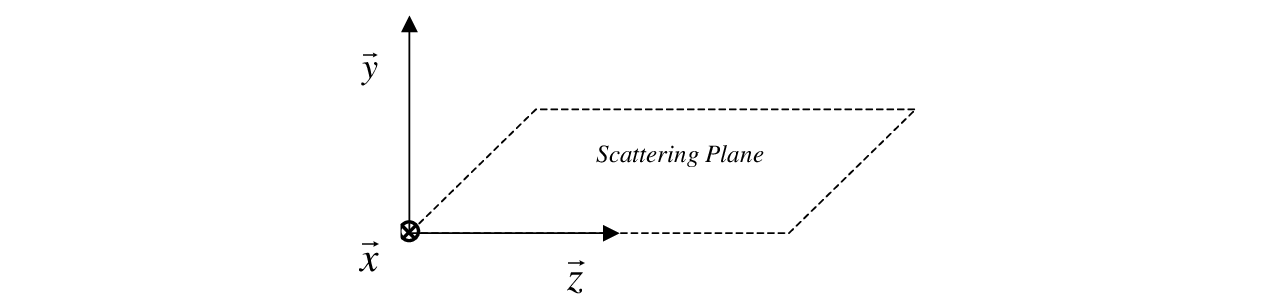, height=1.5in} \caption {The co-ordinate
system. \label{fig:cosys}}
\end{figure}
\begin{figure}
\epsfig{figure=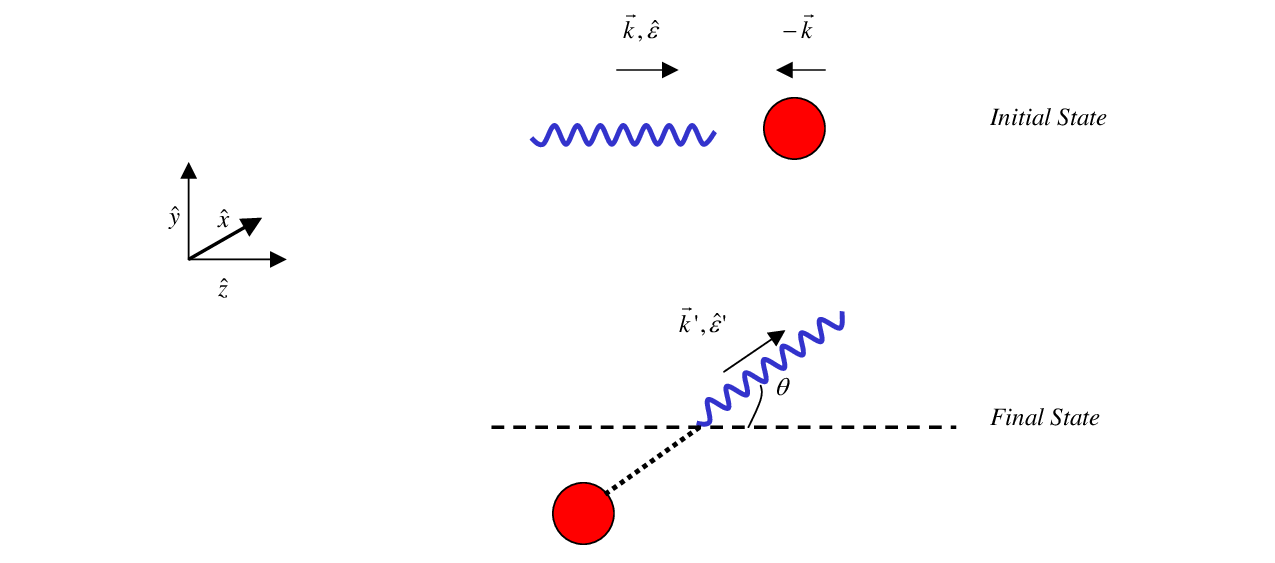, height=2.5in} \caption {Compton
scattering - before and after picture. \label{fig:scat}}
\end{figure}
Before we go further into the discussion of the calculation, let us
first mention that in the subsequent descriptions the following
convention for the co-ordinate system is used (see
Fig.~\ref{fig:cosys}). The beam direction is defined as the $z$
axis. The $x-z$ plane is the scattering plane with the $y$ axis
being normal to it. Fig.~\ref{fig:scat} shows the before and after
picture for Compton scattering from a target. The wiggly line
depicts a photon, the red circle depicts the target. As mentioned
earlier, the scattering takes place in the $x-z$ plane with $\theta$
defining the angle of scattering in the center of mass (c.m.) frame.
For linearly polarized photons, the polarization vectors in the
initial state can be along the $x$ or the $y$ axis, $i.e.$, $\veps =
\hat{x}$ or $\hat{y}$. For circularly polarized photons in the
initial state, the beam helicity, $\la$, can be $\pm1$. The photons
are right circularly polarized (RCP) if the beam helicity, $\la =
+1$ ($\veps_+ = - \frac{\hat{x}+i\hat{y}}{\sqrt{2}}$) or left
circularly polarized (LCP) if the beam helicity, $\la = -1$
($\veps_- = \frac{\hat{x}-i\hat{y}}{\sqrt{2}}$).

\subsection{The Matrix Element}
\label{sec:matel}

Focusing back on our matrix element calculation, we now introduce
complete sets of momentum states to obtain--
\begin{eqnarray}
{\mathcal M}(M_J',M_J) &=& 3 \bra \Psi_f{({1\over 2};M_J')}|{\hat O}(1,2)|\Psi_i{({1\over 2};M_J)} \ket \nonumber \\
&=& 3 \int d^3p_{12}\,' \int d^3p_3\,' \int d^3p_{12} \int d^3p_3 \sum \limits_{\alpha , \alpha', m_{12}, m_{12}'}\nonumber \\
&& \bra \Psi_f{({1\over 2};M_J')}|\vec p_{12}\,' \vec p_3\,'
\alpha'\ket \bra \vec p_{12}\,' \vec p_3\,' \alpha'|{\hat
O}(1,2)|\vec p_{12} \vec p_3 \alpha \ket \bra \vec p_{12} \vec p_3
\alpha
|\Psi_i{({1\over 2};M_J)} \ket \nonumber \\
&=& \int d^3p_{12}\,' \int d^3p_3\,' \int d^3p_{12} \int d^3p_3 \nonumber \\
&& \sum \limits_{\alpha , \alpha', m_{12}, m_{12}'} \varphi (p_{12}\,',p_3\,',\alpha') \varphi (p_{12},p_3,\alpha) \nonumber \\
&& {\mathcal Y}_{l_{12}',s_{12}',j_{12}', m_{12}'}^* (\hat
p_{12}\,') {\mathcal Y}_{l_3',{1\over 2},j_3', M_J'-m_{12}'}^* (\hat
p_3\,') \delta (\vec p_3\,' -(\vec p_3 + {\vec q \over
3})) \nonumber \\
&& \bra \vec p_{12}\,' \vec p_3\,'; \alpha_T'|{\hat O}(1,2)| \vec
p_{12} \vec p_3; \alpha_T \ket {\mathcal Y}_{l_{12},s_{12},j_{12},
m_{12}} (\hat p_{12}) {\mathcal Y}_{l_3,{1\over 2},j_3, M_J-m_{12}}
(\hat p_3). \label{eq9}
\end{eqnarray}
where $|\Psi_i({1\over 2};M_J)\ket$ and $\bra \Psi_f({1\over
2};M_J')|$ denote that the $^3$He nucleus is in a specific isospin
and total spin projected state. In Eq.~(\ref{eq9}), we define
\begin{equation}
{\mathcal Y}_{l,s,j,m}(\hat p) = \sum \limits_{m_s}
(l,m-m_s,s,m_s|l,s,j,m) Y_{l,m-m_s} (\hat p) |s m_s \ket
\label{eq10}
\end{equation}
and
\begin{eqnarray}
\varphi (p_{12},p_3,\alpha) =
(j_{12},m_{12},j_{3},M_J-m_{12}|j_{12},j_{3},J,M_J) \nonumber
\\
(t_{12},mt_{12},{1 \over 2},mt_3|t_{12},{1 \over 2},{1 \over 2},{1
\over 2})\bra p_{12}p_3 \alpha |\Psi \ket.\label{eq10b}
\end{eqnarray}
where the \he3~ wavefunction $\bra p_{12}p_3\al|\Psi \ket$ is
independent of $M_J$.

For a consistent calculation of the matrix element, the nuclear
interaction used to generate the \he3~ wavefunction should be based
on the same chiral effective field theory approach that we used for
the derivation of the operators. Strictly speaking, the available
nuclear interactions have been derived in slightly different
frameworks. Therefore, in order to get an idea of the dependence of
our results on the choice of the wavefunction, we have employed
several nuclear interactions to obtain the \he3~ wavefunctions.
These are the chiral $NLO$ interactions of Refs.~\cite{Ep99, Ep00},
the chiral Idaho $N^3LO$ interaction~\cite{En03} and the high
precision model interaction~\cite{Wi95}. In order to correct for the
underbinding of the \he3~ system when only NN interactions are used,
we augment these interactions also by NNN interactions~\cite{No06,
Pu95} as specified later. Based on this range of interactions, the
binding energy of \he3~ is between 6.89--7.83~MeV (the experimental
one is 7.72 MeV). It turns out that a rather small number of partial
waves is sufficient to sufficient to achieve convergence for the
Compton scattering matrix elements. We found that the one-body
(two-body) matrix elements are converged to within 0.1\% (0.2\%)
using partial waves with $j_{12} \leq$2 ($j_{12} \leq$1).

In the sum in Eq.~(\ref{eq9}), we can use the fact that the isospin
projection of the third particle remains unchanged, as does the
total isospin projection, to introduce the kronecker-delta
$\delta_{mt_{12}', mt_{12}}$. Eq.~(\ref{eq9}) is a 12-dimensional
integral and it can be reduced to a 10-dimensional one by simply
implementing the delta-functions over the angular parts of $\vec
{p_3\,'}$. Since we are separating the angular and magnitude part of
the three-momenta, from here on we shall drop the `vector' sign in
the momenta; for example $|\vec p_3\,'| \equiv p_3\,'$. This leads
to--
\begin{eqnarray}
{\mathcal M}(M_J',M_J) &=& 3 \sum \limits_{j_{12}', j_{12},
m_{12}',m_{12}, s_{12}', s_{12}, l_{12}', l_{12},
j_3', j_3, l_3', l_3,mt_{12}} \nonumber \\
 && \int p_{12}\,'^2 dp_{12}\,'  \int p_{12}^2
dp_{12} \int dp_{3}' \int p_{3}^2 dp_{3} \nonumber \\
&& \varphi (p_{12}\,',p_3\,',\alpha') \varphi (p_{12},p_3,\alpha)
\nonumber \\
&& \int d{\hat p_{12}\,'} \int d{\hat p_{12}} {\mathcal
Y}_{l_{12}',s_{12}',j_{12}', m_{12}'}^* (\hat p_{12}\,')\bra \vec
p_{12}\,'; t_{12}' mt_{12}|{\hat O}(1,2)| \vec p_{12}; t_{12}
mt_{12}\ket  \nonumber
\\
&& {\mathcal Y}_{l_{12},s_{12},j_{12}, m_{12}} (\hat p_{12}) \int
d{\hat p_{3}}{\mathcal Y}_{l_3',{1\over 2},j_3', M_J'-m_{12}'}^*
\left(\widehat {p_3+{q
\over 3}}\right) \nonumber \\
&& \delta \left( p_3\,' -| \vec{p_3} + \vec {{ q \over
3}}|\right){\mathcal Y}_{l_3,{1\over 2},j_3, M_J-m_{12}} (\hat p_3) \nonumber \\
&=& 3 \sum \limits_{j_{12}', j_{12}, m_{12}',m_{12}, s_{12}',
s_{12}, l_{12}', l_{12},
j_3', j_3, l_3', l_3,mt_{12}} \nonumber \\
&& \int p_{12}\,'^2 dp_{12}\,'  \int p_{12}^2
dp_{12} \int p_{3}^2 dp_{3} \varphi (p_{12},p_3,\alpha) \nonumber \\
&& {\mathcal I}_2 (p_{12}, p_{12}\,'; l_{12}, l_{12}', s_{12},
s_{12}', j_{12}, j_{12}', m_{12},
m_{12}', mt_{12})\nonumber \\
&&  \int dp_3\,' \varphi (p_{12}\,',p_3\,',\alpha') {\mathcal
I}_3(p_{3}, p_{3}'; l_{3}, l_{3}', j_{3}, j_{3}', m_{12}, m_{12}',
M_J, M_J') .\label{eq11}
\end{eqnarray}
The integral
\begin{eqnarray} &&{\mathcal I}_2(p_{12}, p_{12}\,';
l_{12}, l_{12}', s_{12}, s_{12}', j_{12}, j_{12}', m_{12}, m_{12}',
mt_{12})= \nonumber \\
&& \int d{\hat p_{12}\,'} \int d{\hat p_{12}}\, {\mathcal
Y}_{l_{12}',s_{12}',j_{12}', m_{12}'}^* (\hat p_{12}\,') \nonumber \\
&&\bra \vec p_{12}\,'; t_{12}' mt_{12}|{\hat O}(1,2)| \vec p_{12};
t_{12} mt_{12}\ket {\mathcal Y}_{l_{12},s_{12},j_{12}, m_{12}} (\hat
p_{12})
\end{eqnarray}
is a four-dimensional integral and is computed numerically. By using
the delta-function we can reduce the dimensionality of
\begin{eqnarray}
&&{\mathcal I}_3(p_{3}, p_{3}'; l_{3}, l_{3}', j_{3}, j_{3}',
m_{12}, m_{12}', M_J, M_J')= \nonumber \\
&& \int d{\hat p_{3}}{\mathcal Y}_{l_3',{1\over 2},j_3',
M_J'-m_{12}'}^* \left(\widehat {p_3+{q \over 3}}\right) \delta
\left( p_3\,' -| \vec{p_3} + \vec {{ q \over 3}}|\right){\mathcal
Y}_{l_3,{1\over 2},j_3, M_J-m_{12}} (\hat p_3)
\end{eqnarray}
which is two-dimensional, thereby making the final integral a
nine-dimensional one. We shall first talk about how ${\mathcal
I}_3(p_{3}, p_{3}'; l_{3}, l_{3}', j_{3}, j_{3}', m_{12}, m_{12}',
M_J, M_J')$ is reduced and then focus on ${\mathcal I}_2(p_{12},
p_{12}\,'; l_{12}, l_{12}', s_{12}, s_{12}', j_{12}, j_{12}',
m_{12}, m_{12}', mt_{12})$. However, even at that point our job is
far from being done because the operators also have to be evaluated
in this same basis. We shall discuss this procedure right after we
finish discussing how the integrals were calculated.

\subsubsection{Reduction of Integral ${\mathcal I}_3$}
\label{sec:I3}

As noted in Eq.~(\ref{eq11}), ${\mathcal I}_3$ is a two-dimensional
integral involving the angular integration of the momentum ${\vec
{p_3}}$. However, we can translate the delta-function in ${\vec
{|p_3\,'}}|$ to one that involves the azimuthal angle of ${\vec
{p_3}}$ using the following steps.
\begin{eqnarray}
\delta \left( p_3\,' -|\vec{p_3} + \vec{q \over 3}|\right) &=&
2p_3\,'\delta \left( p_3\,'^2 -|\vec{p_3} + \vec{q \over
3}|^2\right) \nonumber
\\
&=& 2p_3\,' \delta \left( p_3\,'^2-p_3^2-{q^2 \over 9}-\frac
{2p_3q}{3} (\widehat {p_3 \cdot q}) \right). \label{eq12}
\end{eqnarray}
Since, we work in the co-ordinate system defined in
Sec.~\ref{sec:kin}, ${\hat k}=(0,0,1)$ and ${\hat k'}=(\sin
\theta,0,\cos \theta)$ and this means that ${\hat q}$ is in the
$x-z$ plane or $\phi_q=0$. Then defining ${\hat {p_3}}=(\sin
\theta_3 \cos \phi_3,\sin \theta_3 \sin \phi_3,\cos \theta_3)$ we
can rewrite Eq.~(\ref{eq12}) as--
\begin{eqnarray}
\delta \left( p_3\,' -|\vec{p_3} + \vec{q \over 3}|\right)
&=&\frac{3p_3\,'}{p_3q} \frac{1}{\sin \theta_3 \sin \theta_q} \delta
\left(\cos \phi_3 - \frac{z_3 - \cos \theta_3 \cos \theta_q}{\sin
\theta_3 \sin \theta_q}\right) \nonumber \\
&=& \frac{3p_3\,'}{p_3q} \frac{1}{\sin \theta_3 \sin \theta_q}
\frac{1}{|\sin \tilde{\phi_3}|} \left[\delta (\phi_3 -
\tilde{\phi_3}) + \delta (\phi_3 + \tilde{\phi_3})\right],
\label{eq13}
\end{eqnarray}
where,
\begin{eqnarray}
z_3 &=& \frac{p_3\,'^2-p_3^2-{q^2 \over 9}}{\frac{2p_3q}{3}}
\nonumber \\
\cos \tilde{\phi_3}&=& \frac{z_3-\cos \theta_3 \cos \theta_q}{\sin
\theta_3 \sin \theta_q}. \label{eq14}
\end{eqnarray}
Using Eq.~(\ref{eq13}) we can now eliminate the $\phi_3$ integral in
${\mathcal I}_3$. The delta-function $\delta \left( p_3\,'
-|\vec{p_3} + \vec{q \over 3}|\right)$ further sets the bounds of
integration for both the $p_3\,'$ and the $\theta_3$ integrals.
Implementing the above steps, we can now express the last line of
Eq.~(\ref{eq11}) as--
\begin{eqnarray}
&&\int dp_3\,' \varphi (p_{12}\,',p_3\,',\alpha') {\mathcal
I}_3(p_{3}, p_{3}'; l_{3}, l_{3}', j_{3}, j_{3}', m_{12}, m_{12}',
M_J, M_J')
=\nonumber \\
&& \int \limits_{ |{p_3} - {q \over 3}|}^{{p_3} + {q \over 3}}
dp_3\,' \varphi (p_{12}\,',p_3\,',\alpha')
\frac{3p_3\,'}{p_3 q \sin \theta_q} \nonumber \\
&& \int \limits_{|\theta_q - \arccos z_3|}^{\theta_q + \arccos z_3}
\sin \theta_3 d\theta_3 \frac{1}{|\sin \tilde{\phi_3}| \sin
\theta_3} \sum \limits_{ms_3',ms_3} \delta_{ms_3',ms_3} \nonumber
\\
&& (l_3', M_J'-m_{12}'-ms_3',{1 \over 2}, ms_3'|l_3',{1 \over
2},j_3',M_J'-m_{12}')\nonumber
\\
&&(l_3, M_J-m_{12}-ms_3,{1 \over 2}, ms_3|l_3,{1
\over 2},j_3,M_J-m_{12}) \nonumber \\
&& ( [Y_{l_3',M_J'-m_{12}'-ms_3'}^*(\widehat {p_3 + {q \over
3}})Y_{l_3,M_J-m_{12}-ms_3}(\hat {p_3})]_{\phi_3 = \tilde{\phi_3}}
\nonumber
\\
&& + [Y_{l_3',M_J'-m_{12}'-ms_3'}^*(\widehat {p_3 + {q \over
3}})Y_{l_3,M_J-m_{12}-ms_3}(\hat {p_3})]_{\phi_3 = -\tilde{\phi_3}}
)
\nonumber \\
&=& \int \limits_{ |{p_3} -  {q \over 3}|}^{{p_3} +  {q \over 3}}
dp_3\,' \varphi (p_{12}\,',p_3\,',\alpha') \tilde {{\mathcal I}_3}
(p_{3}, p_{3}'; l_{3}, l_{3}', j_{3}, j_{3}', m_{12}, m_{12}', M_J,
M_J'). \label{eq15}
\end{eqnarray}
with $\tilde {{\mathcal I}_3} (p_{3}, p_{3}'; l_{3}, l_{3}', j_{3},
j_{3}', m_{12}, m_{12}', M_J, M_J')$ as a one-dimensional integral
over $\theta_3$. The kronecker-delta $\delta_{ms_3',ms_3}$ comes
from the fact that our operator does not act on the spin of the
third nucleon. Thus, the final expression for the matrix element
looks like--
\begin{eqnarray}
{\mathcal M}(M_J',M_J) &=& 3 \sum \limits_{j_{12}', j_{12},
m_{12}',m_{12}, s_{12}', s_{12}, l_{12}', l_{12},
j_3', j_3, l_3', l_3,mt_{12}} \nonumber \\
&& \int p_{12}\,'^2 dp_{12}\,'  \int p_{12}^2
dp_{12} \int p_{3}^2 dp_{3} \varphi (p_{12},p_3,\alpha) \nonumber \\
&& {\mathcal I}_2 (p_{12}, p_{12}\,'; l_{12}, l_{12}',
s_{12}, s_{12}', j_{12}, j_{12}', m_{12}, m_{12}', mt_{12})\nonumber \\
&& \int \limits_{ |{p_3} -  {q \over 3}|}^{{p_3} +  {q \over 3}}
dp_3\,'
\varphi (p_{12}\,',p_3\,',\alpha') \nonumber \\
&&\tilde {{\mathcal I}_3}(p_{3}, p_{3}'; l_{3}, l_{3}', j_{3},
j_{3}', m_{12}, m_{12}', M_J, M_J').\label{eq16}
\end{eqnarray}

\subsubsection{Further Reduction of One-Body Matrix Elements}

At this point, we would like to recall that the operator has a
one-body and a two-body part and the kinematics make it further
evident that the one-body calculations should be simpler and involve
fewer integrals. The one-body matrix element is--
\begin{equation}
{\mathcal M} = 3\bra \Psi_f|\frac{1}{2} \big( {\hat O}^{1B}(1)+{\hat
O}^{1B}(2) \big)|\Psi_i\ket = {3 \over 2}(\bra \Psi_f|{\hat
O}^{1B}(1,2) |\Psi_i\ket)
\end{equation}
where ${\hat O}^{1B}(1,2)$ is the one-body operator in the
two-nucleon spin-isospin space. This means that only one of the
three nucleons is involved in the scattering process and we are
gifted with yet another three-momentum conserving delta-function.
For the purpose of this discussion, considering nucleon `1' (see
Fig.~\ref{fig:kin}) as the `struck' nucleon, we can use--
\begin{eqnarray}
-\vec{p}_{12}\,' - {\vec p_3\,' \over 2} - {\vec k' \over 3} &=&
-\vec p_{12} - {\vec p_3 \over 2} - {\vec k \over 3}
\nonumber \\
\Rightarrow \vec p_{12}\,' &=& \vec p_{12} - {\vec q \over 2},
\label{vecp2}
\end{eqnarray}
after using Eq.~(\ref{vecp3}). We have to be careful here because in
our calculations we use the basis of two-nucleon spin and isospin.
However, we use the fact that the spin and isospin sums are
performed only over allowed partial waves and hence we can assume
that nucleon `1' is the struck nucleon and use Eq.~(\ref{vecp2}) to
relate $\vec p_{12}\,' = \vec p_{12} - {\vec q \over 2}$. This is
very useful because we can now reduce integral ${\mathcal I}_2$ in
the same manner as ${\mathcal I}_3$. After going through all the
steps described in Sec.~\ref{sec:I3} we arrive at--
\begin{eqnarray}
{\mathcal M}^{1B}(M_J',M_J) &=& {3 \over 2} \sum \limits_{j_{12}',
j_{12}, m_{12}',m_{12}, s_{12}', s_{12}, l_{12}', l_{12},
j_3', j_3, l_3', l_3,mt_{12}} \nonumber \\
&& \int p_{12}^2 dp_{12} \int p_{3}^2 dp_{3} \varphi (p_{12},p_3,\alpha)  \nonumber \\
&& \int \limits_{|{p_3} - {q \over 3}|}^{{p_3} + {q \over 3}}
dp_3\,' \int \limits_{|{p_{12}} - {q \over 2}|}^{ p_{12} + {q \over
2}}
dp_{12}\,' \varphi (p_{12}\,',p_3\,',\alpha') \nonumber \\
&& \tilde {{\mathcal I}_2} (p_{12}, p_{12}\,'; l_{12}, l_{12}',
s_{12}, s_{12}', j_{12}, j_{12}', m_{12},
m_{12}', mt_{12}) \nonumber \\
&&\tilde {{\mathcal I}_3}(p_{3}, p_{3}'; l_{3}, l_{3}', j_{3},
j_{3}', m_{12}, m_{12}', M_J, M_J').\label{eq17}
\end{eqnarray}
where,
\begin{eqnarray}
&&\tilde {{\mathcal I}_2}  (p_{12}, p_{12}\,'; l_{12}, l_{12}',
s_{12}, s_{12}', j_{12}, j_{12}', m_{12}, m_{12}', mt_{12})\nonumber
\\
&=& \frac{2p_{12}\,'}{p_{12} q \sin \theta_q} \int
\limits_{|\theta_q - \arccos z_{12}|}^{\theta_q +
\arccos z_{12}} \sin \theta_{12} d\theta_{12} \nonumber \\
&& \frac{1}{|\sin \tilde{\phi_{12}}| \sin \theta_{12}} \sum
\limits_{ms_{12}',ms_{12}} \nonumber
\\
&& (l_{12}', m_{12}'-ms_{12}',s_{12}',
ms_{12}'|l_{12}',s_{12}',j_{12}',m_{12}')\nonumber
\\
&&(l_{12}, m_{12}-ms_{12},s_{12},
ms_{12}|l_{12},s_{12},j_{12},m_{12}) \nonumber \\
&& ( [Y_{l_{12}',m_{12}'-ms_{12}'}^*(\widehat {p_{12} - {q \over
2}})Y_{l_{12},m_{12}-ms_{12}}(\hat {p_{12}})]_{\phi_{12} =
\tilde{\phi_{12}}} \nonumber
\\
&& + [Y_{l_{12}',m_{12}'-ms_{12}'}^*(\widehat {p_{12} - {q \over
2}})Y_{l_{12},m_{12}-ms_{12}}(\hat {p_{12}})]_{\phi_{12} =
-\tilde{\phi_{12}}} ) \nonumber \\
&& \bra t_{12}' mt_{12}|\bra s_{12}' ms_{12}'|{\hat O}^{1B}(1,2)|
s_{12} ms_{12}\ket |t_{12} mt_{12}\ket \label{eq18}
\end{eqnarray}
and,
\begin{eqnarray}
z_{12} &=& \frac{p_{12}^2+{q^2 \over 4}-p_{12}\,'^2}{p_{12}q}
\nonumber \\
\cos \tilde{\phi_{12}}&=& \frac{z_{12}-\cos \theta_{12} \cos
\theta_q}{\sin \theta_{12} \sin \theta_q}. \label{eq19}
\end{eqnarray}

Now that we have the structure of the matrix elements, let us look
at the operators. We have to translate the one-body operator (from
App.~\ref{sec:gaN}) and the two-body operators (from
App.~\ref{sec:gaNN}) into the two-nucleon spin-isospin space.

\subsubsection{One-Body operator in Two-Nucleon Spin-Isospin Space}

The one-body operator as described in App.~\ref{sec:gaN} can not be
used directly. So, using Eq.~(\ref{eq:Ti}) we write--
\begin{equation}
{\hat O}^{1B}(1,2) = T_{\gamma N}^{(1)}+T_{\gamma N}^{(2)} = \sum
\limits_{i=1\ldots 6} (A_i^{(1)} t_i^{(1)} + A_i^{(2)} t_i^{(2)})
\label{eq20}
\end{equation}
where $t_i$ are the operators involving the photon polarization,
momenta and nucleon spin and the superscripts `(1)' and `(2)' refer
to specific nucleons. Then defining
\begin{equation}
A_i = A_i^{(IS)} + A_i^{(IV)} \tau_3
\end{equation}
where, the superscripts $(IS)$ and $(IV)$ refer to isoscalar and
isovector pieces of $A_i$ respectively, we can rewrite
Eq.~(\ref{eq20}) as--
\begin{eqnarray}
{\hat O}^{1B}(1,2) &=&\sum \limits_{i=1\ldots
6}[A_i^{(IS)}(t_i^{(1)}+t_i^{(2)})+A_i^{(IV)}(t_i^{(1)}\tau_3^{(1)}+t_i^{(2)}\tau_3^{(2)})]
\nonumber \\
&=&\sum \limits_{i=1\ldots 6}[A_i^{(IS)}(t_i^{(1)}+t_i^{(2)})+{1
\over 2}A_i^{(IV)}\{(\tau_3^{(1)}+\tau_3^{(2)})(t_i^{(1)}+t_i^{(2)})
\nonumber
\\
&&+(\tau_3^{(1)}-\tau_3^{(2)})(t_i^{(1)}-t_i^{(2)})\}]. \label{1Bo}
\end{eqnarray}
And we have,
\begin{equation}
\bra t_{12}'mt_{12}|(\tau_3^{(1)}+\tau_3^{(2)})|t_{12}mt_{12}\ket =
2 \label{t1pt2}
\end{equation}
only for $t_{12}'=t_{12}=1$ and $mt_{12}=1$. (We do not consider the
case when $mt_{12}=-1$ because it is not possible for a \he3~
nucleus.) Also,
\begin{equation}
\bra t_{12}'mt_{12}|(\tau_3^{(1)}-\tau_3^{(2)})|t_{12}mt_{12}\ket =
1
 \label{t1mt2}
\end{equation}
for $\bra 10|\leftarrow |00 \ket$ and $\bra 00|\leftarrow |10 \ket$
only and the rest are zero.

For the spin part of the operator we have either
$(t_i^{(1)}+t_i^{(2)})$ or $(t_i^{(1)}-t_i^{(2)})$. And we can
write--
\begin{equation}
(t_i^{(1)}+t_i^{(2)}) = 2(t_1+t_2) + 2i \sum \limits_{i=3 \ldots
6}\vec S \cdot \vec V_i \label{s1ps2}
\end{equation}
where $\vec S$ is the total spin operator of the two-nucleon system
and $\vec V_i$ is a vector that contains the photon polarization and
momenta. Similarly,
\begin{equation}
(t_i^{(1)}-t_i^{(2)}) = i \sum \limits_{i=3 \ldots 6}(\vec
\sigma^{(1)}-\vec \sigma^{(2)}) \cdot \vec V_i \label{s1ms2}
\end{equation}
This operator induces the transitions $\bra 00|\leftarrow |1 ms_{12}
\ket$ and $\bra 1 ms_{12}'|\leftarrow |00 \ket$.

\subsubsection{The One-Body Matrix Element}

At this point, lets summarize how the different pieces combine to
give ${\mathcal M}^{1B}(M_J',M_J)$. Since, $\tilde {{\mathcal
I}_3}(p_{3}, p_{3}'; l_{3}, l_{3}', j_{3}, j_{3}', m_{12}, m_{12}',
M_J, M_J')$ does not have any pieces of the Compton operator, it is
calculated separately. Then using the information from
Eqs.~(\ref{t1pt2}) and (\ref{t1mt2}) we calculate the isospin matrix
elements in two-body isospin space. Now, for each of the non-zero
isospin transitions, the spin transition matrix elements are
calculated using the information from Eqs.~(\ref{s1ps2}) and
(\ref{s1ms2}), $i.e.$ by calculating the expectation values of $\vec
S$ or $(\vec \sigma^{(1)}-\vec \sigma^{(2)})$ between two-nucleon
spin states as dictated by Eq.~(\ref{1Bo}). Now that we have $\bra
t_{12}' mt_{12}|\bra s_{12}' ms_{12}'|{\hat O}^{1B}(1,2)| s_{12}
ms_{12}\ket |t_{12} mt_{12}\ket $, we put this in Eq.~(\ref{eq18})
and numerically calculate $\tilde {{\mathcal I}_2}  (p_{12},
p_{12}\,'; l_{12}, l_{12}', s_{12}, s_{12}', j_{12},
j_{12}',$\linebreak $m_{12}, m_{12}', mt_{12})$. Then we plug in the
values of $\tilde {{\mathcal I}_3}(p_{3}, p_{3}'; l_{3}, l_{3}',
j_{3}, j_{3}', m_{12}, m_{12}', M_J, M_J')$ and $\tilde {{\mathcal
I}_2} (p_{12}, p_{12}\,'; l_{12}, l_{12}', s_{12}, s_{12}', j_{12},
j_{12}', m_{12}, m_{12}', mt_{12})$ in Eq.~(\ref{eq17}), evaluate
the integrals over $p_3\,'$, $p_3$, $p_{12}\,'$ and $p_{12}$ after
introducing the wavefunctions projected into the equivalent basis
and finally, sum over all angular momentum and isospin quantum
numbers to obtain ${\mathcal M}^{1B}(M_J',M_J)$.

Next, we shall describe the procedure for the two-body operators and
put everything back together to obtain the two-body matrix element.

\subsubsection{Structure of Two-Body Matrix Elements}

The two-body matrix element can be written as--
\begin{eqnarray}
{\mathcal M}^{2B}(M_J',M_J) &=& 3 \sum \limits_{j_{12}', j_{12},
m_{12}',m_{12}, s_{12}', s_{12}, l_{12}', l_{12},
j_3', j_3, l_3', l_3,mt_{12}} \nonumber \\
&& \int p_{12}^2 dp_{12} \int p_{12}\,'^2 dp_{12}\,' \int p_{3}^2 dp_{3} \varphi (p_{12},p_3,\alpha)  \nonumber \\
&& \int \limits_{|{p_3} - {q \over 3}|}^{{p_3} + {q \over 3}}
dp_3\,'
\varphi (p_{12}\,',p_3\,',\alpha') \nonumber \\
&& {\mathcal I}_2 (p_{12}, p_{12}\,'; l_{12}, l_{12}', s_{12},
s_{12}', j_{12}, j_{12}', m_{12},
m_{12}', mt_{12}) \nonumber \\
&&\tilde {{\mathcal I}_3}(p_{3}, p_{3}'; l_{3}, l_{3}', j_{3},
j_{3}', m_{12}, m_{12}', M_J, M_J').\label{2bme}
\end{eqnarray}
where, ${\mathcal I}_2(p_{12}, p_{12}\,'; l_{12}, l_{12}', s_{12},
s_{12}', j_{12}, j_{12}', m_{12}, m_{12}', mt_{12})$ in its entire
glory is given by--
\begin{eqnarray}
&&{\mathcal I}_2(p_{12}, p_{12}\,'; l_{12}, l_{12}', s_{12},
s_{12}',
j_{12}, j_{12}', m_{12}, m_{12}', mt_{12})= \nonumber \\
&&\int d{\hat p_{12}\,'} \int {\hat p_{12}} {\mathcal
Y}_{l_{12}',s_{12}',j_{12}', m_{12}'}^* (\hat p_{12}\,') \nonumber \\
&& \bra \vec p_{12}\,'; t_{12}' mt_{12}|{\hat O}(1,2)| \vec p_{12};
t_{12} mt_{12}\ket {\mathcal Y}_{l_{12},s_{12},j_{12}, m_{12}}
(\hat p_{12}) \nonumber \\
&& = \int d{\hat p_{12}\,'} \int {\hat p_{12}} \sum
\limits_{ms_{12},
m_{12}'} (l_{12}',m_{12}'-ms_{12}',s_{12}',ms_{12}'|l_{12}',s_{12}',j_{12}',m_{12}') \nonumber \\
&&(l_{12},m_{12}-ms_{12},s_{12},ms_{12}|l_{12},s_{12},j_{12},m_{12}) \nonumber \\
&& Y^*_{l_{12}',m_{12}'-ms_{12}'}(\hat
p_{12}\,')Y_{l_{12},m_{12}-ms_{12}}(\hat p_{12}) \nonumber \\
&& \bra \vec p_{12}\,'; t_{12}' mt_{12} s_{12}' ms_{12}'|{\hat
O}^{2B}(1,2)|\vec p_{12}; t_{12} mt_{12} s_{12} ms_{12} \ket
\label{2bi2}
\end{eqnarray}
Let us then see how $\bra \vec p_{12}\,'; t_{12}' mt_{12}' s_{12}'
ms_{12}'|{\hat O}^{2B}(1,2)|\vec p_{12}; t_{12} mt_{12} s_{12}
ms_{12} \ket$ is evaluated.

\subsubsection{Two-Body Operator in Two-Nucleon Spin-Isospin Space}

The two-body operator is given by Eq.~(\ref{eq:total}) and each of
the terms in that equation are expressed in the equations following
it. Symbolically, we can write each of those diagrams as (for
example, for the first diagram in Fig.~\ref{fig:2B})--
\begin{eqnarray}
T_{\gamma NN}^{(a)}&= & -\frac{{e^2}{g_A^2}}{2 f_\pi^2} \;
({\vec\tau}^{\; (1)} \cdot{\vec\tau}^{\;
(2)}-\tau^{(1)}_{3}\tau^{(2)}_{3})\frac{{\veps\cdot\vsigone}\;{\vepsprime\cdot\vsigtwo}}
{2\lbrack {\omega^2}-{m_\pi^2}- (\vpee -\vpeeprime
+{\frac{1}{2}(\vkay +\vkayprime )})^2 \rbrack} \nonumber \\
&=&\digamma^{(a)} ({\vec\tau}^{(1)}
\cdot{\vec\tau}^{(2)}-\tau^{(1)}_{3}\tau^{(2)}_{3}){\veps\cdot\vsigone}\;{\vepsprime\cdot\vsigtwo}
\end{eqnarray}
In general, the spin-isospin structure of the two-body operator
would be--
\begin{equation}
\digamma^{(..)} ({\vec\tau}^{(1)}
\cdot{\vec\tau}^{(2)}-\tau^{(1)}_{3}\tau^{(2)}_{3}){\vsigone
\cdot\vec A}\;{\vsigtwo \cdot \vec B}.
\end{equation}
The only transitions in isospin space that are non-zero are $\bra
00|\leftarrow |00 \ket$ and $\bra 10|\leftarrow |1 0 \ket$ and both
of these are--
\begin{equation}
 \bra 00|({\vec\tau}^{\; (1)}
\cdot{\vec\tau}^{\; (2)}-\tau^{(1)}_{3}\tau^{(2)}_{3})|00\ket=-\bra
10|({\vec\tau}^{\; (1)} \cdot{\vec\tau}^{\;
(2)}-\tau^{(1)}_{3}\tau^{(2)}_{3})|10\ket=-2. \label{2biso}
\end{equation}
For the spin part of the operator we can write--
\begin{equation}
{\vsigone \cdot\vec A}\;{\vsigtwo \cdot \vec B} = {1 \over 2} ({\hat
O}_s + {\hat O}_a) \label{eq21}
\end{equation}
where ${\hat O}_s$ is the symmetric combination of the LHS of
Eq.~(\ref{eq21}) and ${\hat O}_a$ is the asymmetric combination
under the interchange of nucleons. Then, we can write ${\hat O}_s$
as--
\begin{equation}
{\hat O}_s = 4 \vec S \cdot \vec A \vec S \cdot \vec B - 2 \vec A
\cdot \vec B -2i \vec S \cdot (\vec A \times \vec B).
\end{equation}
Here, $2\vec S= \vsigone + \vsigtwo$  and we have used the identity
${\sigma \cdot\vec A}\;{\sigma \cdot \vec B} = \vec A \cdot \vec B +
i \vec \sigma \cdot (\vec A \times \vec B)$. Since this is a
spin-symmetric combination ${\hat O}_s$ induces the transitions
$\bra 00|\leftarrow |00 \ket$ and $\bra 1 ms_{12}'|\leftarrow |1
ms_{12} \ket$. The different spin transition matrix elements can be
written in terms of general vectors $\vec A$ and $\vec B$ as --
\begin{eqnarray}
\bra 00|{\hat O}_s|00\ket &=& -2 \vec A \cdot \vec B \nonumber \\
\bra 1 ms_{12}'|{\hat O}_s|1 ms_{12}\ket &=& -2 \vec A \cdot \vec B
\delta_{ms_{12}', ms_{12}} \nonumber \\
&& + 8 \sum \limits_{\mu \alpha \beta} (1, \alpha, 1, \mu |1,1,1,
ms_{12}')(1, \beta, 1, ms_{12}| 1,1,1,
\mu)A_{\alpha}^{\dagger}B_{\beta}^{\dagger} \nonumber \\
&& + i2 \sqrt{2} \sum \limits_{\gamma} (1, \ga, 1, ms_{12} |1,1,1,
ms_{12}') (\vec A \times \vec B)_{\ga}^{\dagger} \label{2bos}
\end{eqnarray}
where we have introduced spherical components of vectors $\vec A$
and $\vec B$. Similarly, we can simplify the asymmetric combination
to--
\begin{equation}
{\hat O}_a = 2 \vec S \cdot \vec A (\vsigone - \vsigtwo) \cdot \vec
B  -i (\vsigone - \vsigtwo) \cdot (\vec A \times \vec B).
\end{equation}
This, being a spin-asymmetric combination, it induces transitions
between $\bra 00|\leftarrow |1 ms_{12} \ket$ or $\bra 1
ms_{12}'|\leftarrow |00 \ket$. After going through some spin algebra
we obtain--
\begin{eqnarray}
\bra 1 ms_{12}'|{\hat O}_a|00\ket &=& 2\sqrt{2}\sum \limits_{\alpha
\beta}(1, \alpha, 1, \be |1,1,1,
ms_{12}')A_{\alpha}^{\dagger}B_{\beta}^{\dagger} \nonumber
\\
&& -i \sum \limits_{\gamma} \delta_{ms_{12}', \ga} (\vec A \times
\vec B)_{\ga}^{\dagger}. \label{2boa}
\end{eqnarray}
We can calculate $\bra 00|{\hat O}_a |1 ms_{12} \ket$ by taking the
hermitian conjugate of Eq.~(\ref{2boa}).

\subsubsection{The Two-Body Matrix Element}

We can now summarize how the different pieces combine to give
${\mathcal M}^{2B}(M_J',M_J)$. Again, $\tilde {{\mathcal
I}_3}(p_{3}, p_{3}'; l_{3}, l_{3}', j_{3}, j_{3}', m_{12}, m_{12}',
M_J, M_J')$ does not have any pieces of the Compton operator, it is
calculated separately. The isospin transition matrix elements are
simple in this case and given by Eq.~(\ref{2biso}). For each of
these two isospin transitions, the spin transition matrix elements
are calculated using the expressions from Eqs.~(\ref{2bos}) and
(\ref{2boa}). (Remember, the vectors $\vec A$ and $\vec B$ are
different combinations of photon polarization and momentum vectors
and nucleon momenta.) Now that we have $\bra \vec p_{12}\,'; t_{12}'
mt_{12} s_{12}' ms_{12}'|{\hat O}^{2B}(1,2)$\linebreak $|\vec
p_{12}; t_{12} mt_{12} s_{12} ms_{12} \ket$, we put this in
Eq.~(\ref{2bi2}) and numerically calculate it. Then we plug in the
values of $\tilde {{\mathcal I}_3}(p_{3}, p_{3}'; l_{3}, l_{3}',
j_{3}, j_{3}', m_{12}, m_{12}', M_J, M_J')$ and ${\mathcal I}_2
(p_{12}, p_{12}\,'; l_{12}, l_{12}', s_{12}, s_{12}', j_{12},
j_{12}',$\linebreak $ m_{12}, m_{12}', mt_{12})$ in
Eq.~(\ref{2bme}), numerically evaluate the integrals over $p_3\,'$,
$p_3$, $p_{12}\,'$ and $p_{12}$ after introducing the wavefunctions
projected into the equivalent basis and finally, sum over all
angular momentum and isospin quantum numbers to obtain ${\mathcal
M}^{2B}(M_J',M_J)$ (Eq.~({\ref{2bme}})).

Finally, we add the one-body and the two-body pieces to obtain--
\begin{equation}
{\mathcal M}(M_J',M_J)={\mathcal M}^{1B}(M_J',M_J)+{\mathcal
M}^{2B}(M_J',M_J)
\end{equation}
and using this we can now calculate various observables. In the next
section we discuss the observables that we concentrate on and then
in Sec.~\ref{sec:res} we discuss some of our results for those
observables.

\section{Observables}
\label{sec:obs}

\subsection{Differential Cross-section}

The differential cross-section (dcs) is proportional to the
spin-averaged sum of the square of the scattering amplitude:
\begin{equation}
\frac{d\sigma}{d\Omega} = \frac{1}{4} \left(\frac{M_{^3He}}{4 \pi\,
\sqrt {s(\w)}}\right)^2 \sum_{M_J', M_J, \vepsprime =
(\hat{x},\hat{y}), \veps = (\hat{x},\hat{y})} |\bra M_J', \vepsprime
| {\hat O} | \veps, M_J \ket|^2. \label{eq:matdcs}
\end{equation}
Here, $M_J$ and $M_J'$ are the initial and final spin projections of
the nucleus respectively, $M_{^3He}$ is the mass of the \he3~
nucleus and $s(\w)$ is Mandelstam $s$. The factor of $\frac{1}{4}$
comes from the averaging over the initial \he3~ spin and photon
polarization states.

\subsection{Double-Polarization Asymmetry}

\begin{figure}
\epsfig{figure=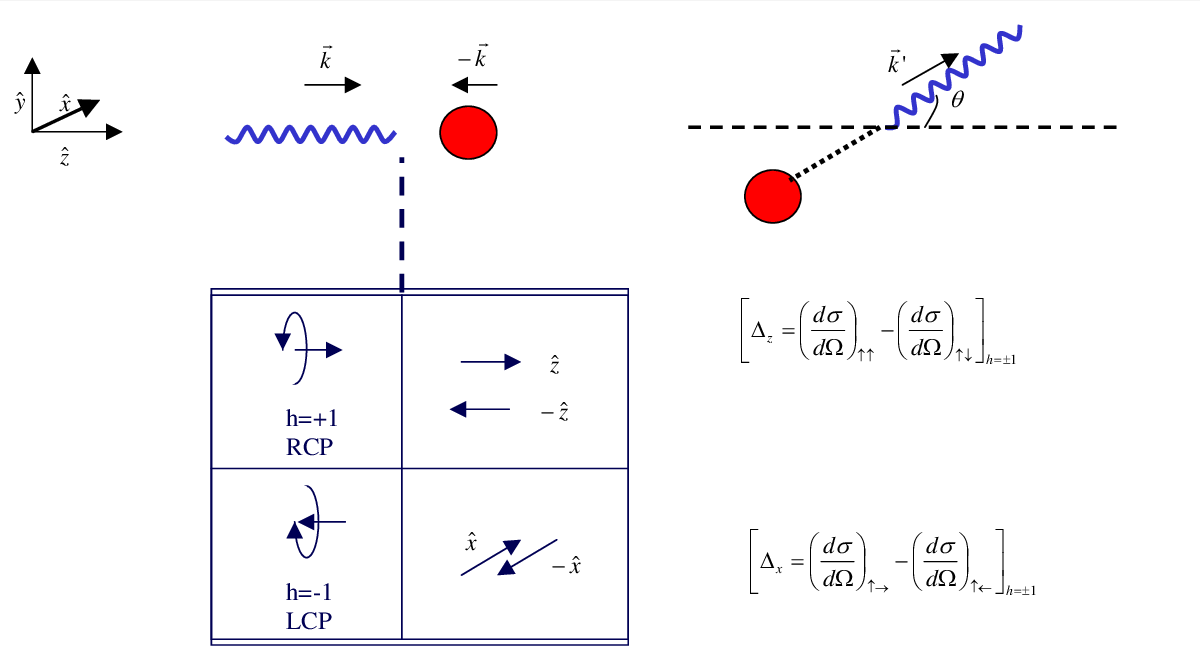, height=3in} \caption{The double-polarization
observables. The inset in the bottom left shows possible circular
polarization states for the incoming photons (on the left) and
possible initial target spin polarized states (on the right).
\label{fig:obs}}
\end{figure}
The double-polarization asymmetry involves circularly polarized
photons and a spin-polarized target (see Fig.~\ref{fig:obs}). The
expectation is that these observables can provide insight into the
spin polarizabilities of the target.

When the target is polarized along the beam direction (parallel or
anti-parallel), the corresponding observable is called the parallel
target polarization asymmetry and is defined to be:
\begin{equation}
\Sigma_{z,(\la = \pm1)} =\frac{(\frac{d\sigma}{d\Omega})_{\uparrow
\uparrow} - (\frac{d\sigma}{d\Omega})_{\uparrow
\downarrow}}{(\frac{d\sigma}{d\Omega})_ {\uparrow \uparrow} +
(\frac{d\sigma}{d\Omega})_{\uparrow \downarrow}}. \label{eq:sigmaz}
\end{equation}
Parallel (anti-parallel) arrows in the subscript symbolize target
polarization parallel (anti-parallel) to the beam helicity and $\la$
denotes the helicity of the incoming photon. This observable can be
depicted only in terms of the numerator of Eq.(\ref{eq:sigmaz}) as a
difference in cross-section as--
\begin{equation}
\Delta_{z,(\la = \pm1)}
=\left(\frac{d\sigma}{d\Omega}\right)_{\uparrow \uparrow} -
\left(\frac{d\sigma}{d\Omega}\right)_{\uparrow \downarrow}.
\label{eq:deltaz}
\end{equation}

The target may be polarized along $\pm \hat{x}$ too. In this case,
the observable is called the perpendicular polarization asymmetry
and is given by:
\begin{equation}
\Sigma_{x,(\la = \pm1)}=\frac{(\frac{d\sigma}{d\Omega})_{\uparrow
\rightarrow} - (\frac{d\sigma}{d\Omega})_{\uparrow
\leftarrow}}{(\frac{d\sigma}{d\Omega})_ {\uparrow \rightarrow} +
(\frac{d\sigma}{d\Omega})_{\uparrow \leftarrow}}. \label{eq:sigmax}
\end{equation}
Again, the direction of the second arrow in the subscript denotes
the target polarization is along the $+\hat{x}$ or $-\hat{x}$
directions. $\la$ is the helicity of the incoming photon. For this
observable one must be careful in defining the matrix elements
because the spin state of the target is an eigenstate of $S_x$ and
not of $S_z$. The eigenstates of $S_x$ should be first expressed in
terms of those of $S_z$ and then one can proceed with the
calculations. As before, this observable can also be expressed as a
difference in cross-sections as--
\begin{equation}
\Delta_{x,(\la =
\pm1)}=\left(\frac{d\sigma}{d\Omega}\right)_{\uparrow \rightarrow} -
\left(\frac{d\sigma}{d\Omega}\right)_{\uparrow \leftarrow}.
\label{eq:deltax}
\end{equation}

It should be mentioned here that the results reported in this paper
are for $\Delta_z$ and $\Delta_x$.

\section{How do Polarizabilities Enter into the Calculation?}
\label{sec:Aw}

The one-body amplitude (App.~\ref{sec:gaN}) has six invariant
structures $A_1 \ldots A_6$ that depend on the photon energy. When
these amplitudes are Taylor-expanded in $\w$ around $\w \sim 0$ we
obtain the following expressions~\cite{McG01}--
\begin{eqnarray}
A_1 &=& -\frac{{\mathcal Z}^2}{M}
        +(\alpha + \beta \cos(\theta))\omega^2 + {\mathcal O}(\w^4),
\nonumber \\
A_2 &=& \frac{{\mathcal Z}^2\omega}{M^2}
        + \beta \omega^2 + {\mathcal O}(\w^4),
\nonumber \\
A_3 &=& \frac{\omega}{2M^2} [{\mathcal Z}({\mathcal Z}+2\kappa
)-({\mathcal Z}+\kappa)^2 \cos \theta] + A_3^{\pi^0} +
        (\ga_1 - (\ga_2 + 2\ga_4)\cos(\theta))\w^3
        +{\mathcal O}(\omega^5),
\nonumber \\
A_4 &=& -\frac{({\mathcal Z}+\kappa )^2\omega}{2M^2} + \ga_2 \w^3
        +{\mathcal O}(\omega^5),
\nonumber \\
A_5 &=& \frac{({\mathcal Z}+\kappa )^2\omega}{2M^2} +  A_5^{\pi^0} +
\ga_4 \w^3
        +{\mathcal O}(\omega^5),
\nonumber \\
A_6 &=& -\frac{{\mathcal Z}({\mathcal Z}+\kappa )\omega}{2M^2} +
A_6^{\pi^0} + \ga_3 \w^3  +{\mathcal O}(\omega^5). \label{eq:Asinw}
\end{eqnarray}
Here, $A_3^{\pi^0}$, $A_5^{\pi^0}$ and $A_6^{\pi^0}$ are
contributions from the $\pi^0$-pole graph (see Fig.~\ref{fig:tree},
lower-most diagram). The zeroth-order term in $\w$ is the
leading-order Thomson term (as in Eq.~(\ref{eq:1})) and the
first-order term contains contributions from the anomalous magnetic
moment. The $\w^2$ term depends on $\al$ and $\be$ and the spin
polarizabilities ($\ga$'s), appear in the $\w^3$ term. At ${\mathcal
O}(e^2 Q)$ the following results for the polarizabilities can be
obtained~\citep{Be91, Be92} by matching Eq.~(\ref{eq:As}) to
Eq.~(\ref{eq:Asinw}):
\begin{eqnarray}
\alpha_p=\alpha_n=\frac{5 e^2 g_A^2}{384 \pi^2 f_\pi^2 m_\pi}
&=&12.2 \times
10^{-4} \, {\rm fm}^3, \label{eq:alphaOQ3} \nonumber \\
\beta_p=\beta_n=\frac{e^2 g_A^2}{768 \pi^2 f_\pi^2 m_\pi}&=& 1.2
\times
10^{-4} \, {\rm fm}^3, \label{eq:betaOQ3} \nonumber \\
\ga_{1p}=\ga_{1n}=\frac{e^2 g_A^2}{98 \pi^3 f_\pi^2 m_\pi^2} &=&4.4
\times
10^{-4} \, {\rm fm}^4, \label{eq:gamma1OQ3} \nonumber \\
\ga_{2p}=\ga_{2n}=\frac{e^2 g_A^2}{192 \pi^3 f_\pi^2 m_\pi^2} &=&2.2
\times
10^{-4} \, {\rm fm}^4, \label{eq:gamma2OQ3} \nonumber \\
\ga_{3p}=\ga_{3n}=\frac{e^2 g_A^2}{384 \pi^3 f_\pi^2 m_\pi^2} &=&1.1
\times
10^{-4} \, {\rm fm}^4, \label{eq:gamma3OQ3} \nonumber \\
\ga_{4p}=\ga_{4n}=-\frac{e^2 g_A^2}{384 \pi^3 f_\pi^2 m_\pi^2}
&=&-1.1 \times 10^{-4} \, {\rm fm}^3. \label{eq:gamma4OQ3}
\end{eqnarray}

The manner in which we investigate the impact of the neutron
polarizabilities is by varying them when calculating any observable
around these central ${\mathcal O}(e^2 Q)$ values. We introduce six
new parameters as corrections to the ${\mathcal O}(e^2 Q)$ values,
which we call $\De \al_n$, $\De \be_n$ and $\De \ga_{in}
(i=1,2,3,4)$:
\begin{eqnarray}
A_1 &=& A_1({\mathcal O}(Q^3)) + (\Delta \alpha_n + \Delta \beta_n
\cos(\theta))\omega^2,
\nonumber \\
A_2 &=& A_2({\mathcal O}(Q^3)) + \Delta \beta_n \omega^2,
\nonumber \\
A_3 &=& A_3({\mathcal O}(Q^3)) +(\Delta \ga_{1n} - (\Delta \ga_{2n}
+ 2\Delta \ga_{4n})\cos(\theta))\w^3,
\nonumber \\
A_4 &=& A_4({\mathcal O}(Q^3))+\Delta \ga_{2n} \w^3,
\nonumber \\
A_5 &=& A_5({\mathcal O}(Q^3))+\Delta \ga_{4n} \w^3,
\nonumber \\
A_6 &=& A_6({\mathcal O}(Q^3)) +\Delta \ga_{3n} \w^3.
\label{eq:AsinDw}
\end{eqnarray}
But, note that, this only includes one set of higher-order
contributions. It does not represent a full calculation at
${\mathcal O}(e^2 Q^2)$ and it only examines a limited set of the
contributions generated by the $\Delta$-isobar.

For one particular plot only one of these parameters, for instance,
$\De \al_n$ is varied with the rest being set arbitrarily to zero.
This gives us a measure of the sensitivity of that particular
observable to $\De \al_n$. In principle, one should fit these
parameters to experimental data and thus extract them, but at
present we are still waiting for data on most of these observables.

\section{Results}
\label{sec:res}

\subsection{Coherent Compton Scattering}

\begin{figure}[!htb]
\begin{center}
\includegraphics*[width=.48\linewidth]{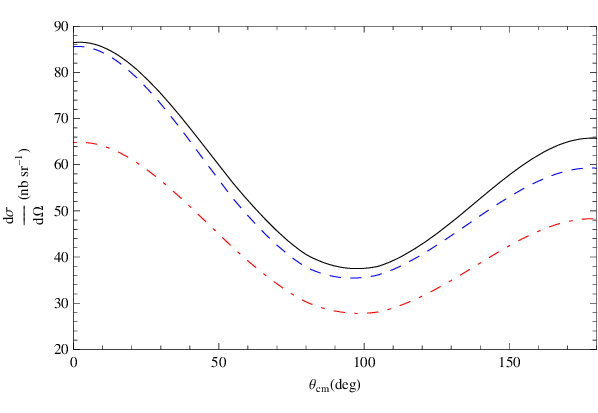}
\hfill
\includegraphics*[width=.48\linewidth]{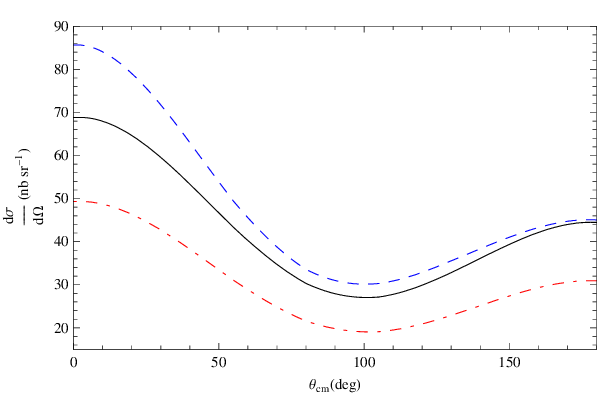}\\
\includegraphics*[width=.48\linewidth]{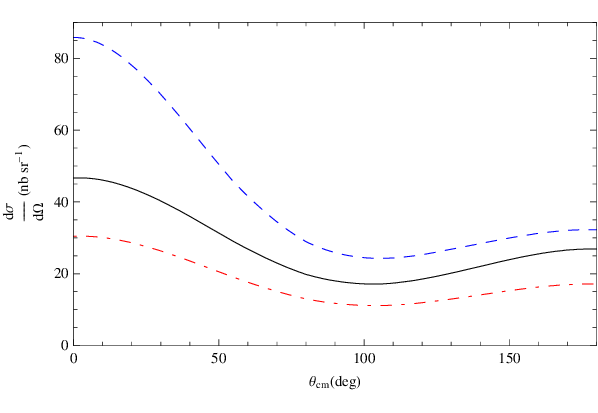}
\hfill
\includegraphics*[bb= 0 0 287 198, width=.48\linewidth]{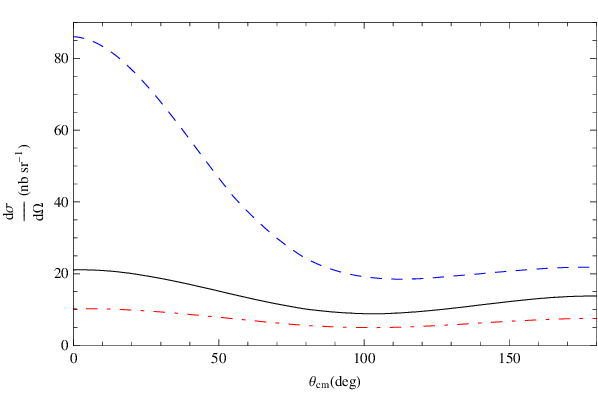}
\parbox{1.\textwidth}{
  \caption {Comparison of differential
cross-section calculated at different orders in the c.m. frame. The
top two panels are for calculations at 60~MeV (left) and 80~MeV
(right) and the bottom two panels are calculations at 100~MeV (left)
and 120~MeV (right). The dashed (blue) curves are the ${\mathcal O}
(e^2)$ results, the solid (black) curves are the impulse
approximation results and the dot-dashed (red) curves are the
${\mathcal O} (e^2 Q)$ results.}
  \label{fig:dcs}}
\end{center}
\end{figure}

In Fig.~\ref{fig:dcs} we plot the dcs for coherent $\ga \, ^3$He
scattering. For the results reported in this section we employ the
\he3~ wavefunction obtained using the Idaho chiral $N^3LO$ NN
potential~\cite{En03} with the cut-off at 500~MeV together with a
chiral three-nucleon interaction~\cite{No06}. All the different
panels are for calculations at different energies (60, 80, 100 and
120 MeV) in the c.m. frame. Each of the panels shows the dcs
calculation at different orders-- $\calO(e^2)$, Impulse
Approximation (IA) and $\calO(e^2 Q)$. Here, Impulse Approximation
means that the calculation is done up to ${\mathcal O}(e^2 Q)$ but
does not have any two-body contribution-- this is not the full
$\calO(e^2 Q)$ calculation. As expected, we see that there is a
sizeable difference between the IA and the ${\mathcal O}(e^2 Q)$ dcs
which means that the two-body currents are important and cannot be
neglected. Also, we see that the difference between the $\calO(e^2)$
and $\calO(e^2 Q)$ is very small at 60 MeV and it gradually
increases with energy. This can be attributed to the fact that, as
energy increases, the energy-dependent contributions from the $A_1
\ldots A_6$ terms increase and this means that there is an
opportunity to extract the neutron polarizabilities. At the lowest
energy the dcs is still dominated by the proton Thomson term and
hence the $\calO(e^2 Q)$ dcs is closest to the $\calO(e^2)$ one, or
in other words, \cpt converges well there. Another notable feature
is that the $\calO(e^2)$ dcs is independent of the c.m. energy at
forward angles, or we can say that the dcs at $\calO(e^2)$ converges
to a single value at zero momentum transfer.

Now, focusing just on the $\calO(e^2 Q)$ curves for all four
energies it is evident that there is a rapid decrease in the dcs as
one goes from 60 MeV to 120 MeV. It remains to be seen whether this
drop-off is compensated by addition of the $\Delta$-isobar to the
theory. Also notable is the fact that the ~\he3 Compton scattering
dcs is around four times bigger in magnitude compared to deuteron
Compton scattering dcs. We conclude that Compton scattering
experiments on \he3~ are potentially more accurate than on deuteron.

\begin{figure}[!htb]
\begin{center}
\includegraphics*[width=.48\linewidth]{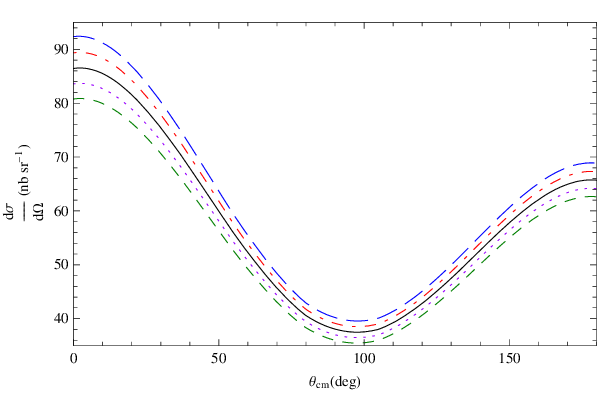}
\hfill
\includegraphics*[width=.48\linewidth]{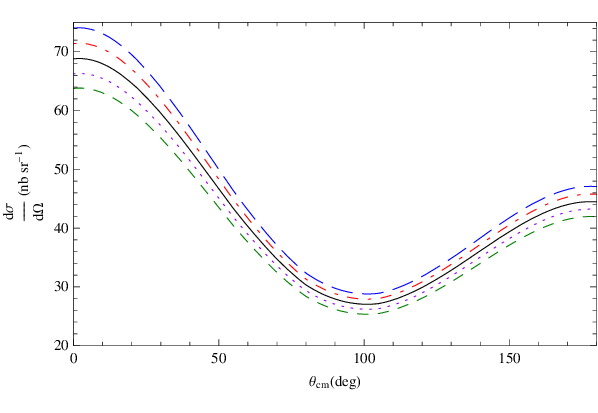}\\
\includegraphics*[width=.48\linewidth]{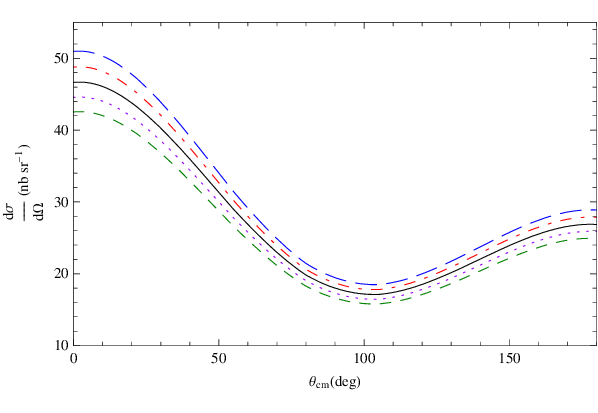}
\hfill
\includegraphics*[width=.48\linewidth]{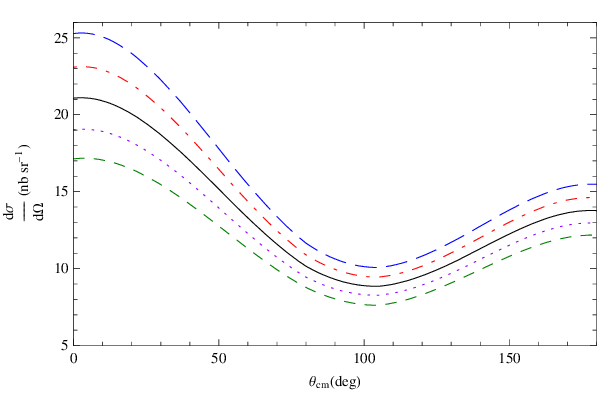}
\parbox{1.\textwidth}{
  \caption {The differential
cross-sections  in the c.m. frame with varying $\Delta \alpha_n$.
The top two panels are for calculations at 60~MeV (left) and 80~MeV
(right) and the bottom two panels are calculations at 100~MeV (left)
and 120~MeV (right). The solid (black) curves correspond to the full
$\calO(e^2 Q)$ results. The long-dashed (blue) curves correspond to
$\Delta \alpha_n = -4 \times 10^{-4}$fm$^3$, dot-dashed (red) to
$\Delta \alpha_n = -2 \times 10^{-4}$fm$^3$, dotted (magenta) to
$\Delta \alpha_n = 2 \times 10^{-4}$fm$^3$ and dashed (green) to
$\Delta \alpha_n = 4 \times 10^{-4}$fm$^3$}
  \label{fig:dcsalpha}}
\end{center}
\end{figure}

\begin{figure}[!htb]
\begin{center}
\includegraphics*[width=.48\linewidth]{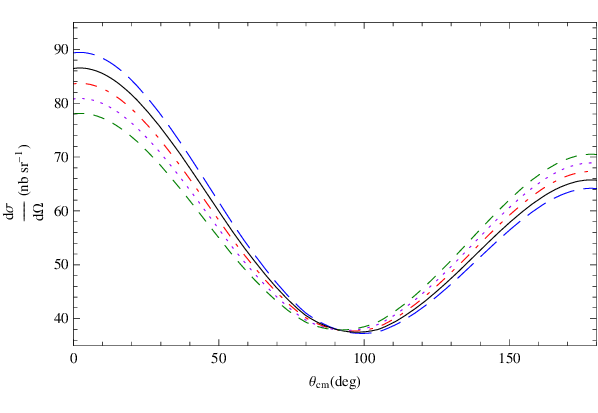}
\hfill
\includegraphics*[width=.48\linewidth]{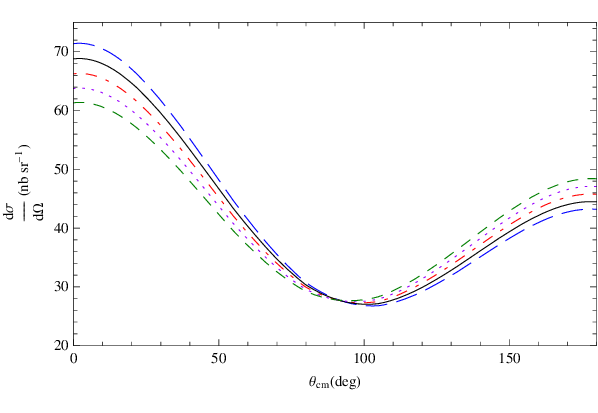}\\
\includegraphics*[width=.48\linewidth]{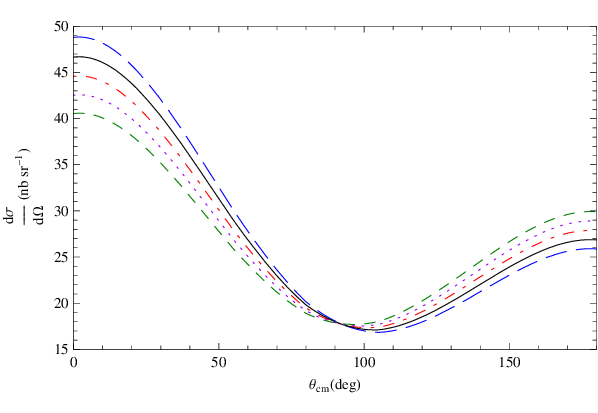}
\hfill
\includegraphics*[width=.48\linewidth]{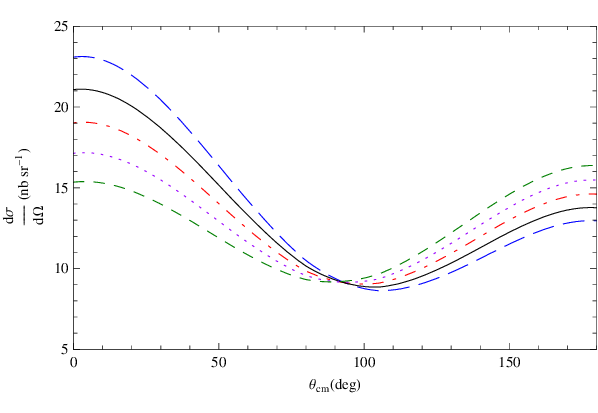}
\parbox{1.\textwidth}{
  \caption {The differential
cross-sections  in the c.m. frame with varying $\Delta \be_n$. The
top two panels are for calculations at 60~MeV (left) and 80~MeV
(right) and the bottom two panels are calculations at 100~MeV (left)
and 120~MeV (right). The solid (black) curves correspond to the full
$\calO(e^2 Q)$ results. The long-dashed (blue) curves correspond to
$\Delta \be_n = -2 \times 10^{-4}$fm$^3$, dot-dashed (red) to
$\Delta \be_n = 2 \times 10^{-4}$fm$^3$, dotted (magenta) to $\Delta
\be_n = 4 \times 10^{-4}$fm$^3$ and dashed (green) to $\Delta \be_n
= 6 \times 10^{-4}$fm$^3$}
  \label{fig:dcsbeta}}
\end{center}
\end{figure}

Next in Figs.~\ref{fig:dcsalpha} and \ref{fig:dcsbeta} we plot the
$\calO(e^2 Q)$ dcs at different energies by varying $\Delta \al_n$
and $\Delta \be_n $ respectively. Here we vary $\Delta \al_n $
between $(-4 \ldots 4) \times 10^{-4}$~fm$^3$ and $\Delta \be_n $
between $(-2 \ldots 6) \times 10^{-4}$~fm$^3$. (The range of
variation of $\Delta \be_n $ is chosen in this manner because of a
strong paramagnetic contribution from the
$\Delta$-isobar~\cite{Hi05a, Hi05}.) What is striking in these
figures is that because the \he3~ dcs is large compared to that of
the deuteron, the sensitivity to $\Delta \al_n $ and $\Delta \be_n $
is also similarly magnified. It should be evident from the figures
that the size of the absolute differences in the dcs as we vary
$\Delta \al_n$ and $\Delta \be_n$ is roughly the same at all
energies. However, we would like to advocate that if such
measurements were necessary (we expect that $\al_n$ and $\be_n$
could be determined from ongoing coherent $\ga d$ experiments at
MAXLab at Lund) then it would be better to do them at lowest energy
possible. At the lower energies, the dominant contributions come
from $\al_n$ and $\be_n$, apart from the proton Thomson terms. In
this regime, we best understand the dynamics and the extraction will
not be tainted by contributions from the spin polarizabilities
entering via $A_3\ldots A_6$ and other higher-order contributions.
The figures show that even at 60 MeV, the sensitivity to $\al_n$ and
$\be_n$ is sizeable. Another message to take away from these two
figures is the beautiful feature in Fig.~\ref{fig:dcsbeta} that
sensitivity to ${\be_n}$ vanishes at $\theta = 90$~deg and the
curves themselves turn over. Thus, in principle one could extract
${\al_n}$ and ${\be_n}$ independently from the same experiment as
opposed to measuring one and then using the sum rule to get the
other. A measurement near $\theta=90$~deg would yield ${\al_n}$ and
then one could perform a forward/backward dcs ratio measurement to
extract ${\be_n}$. The various curves for $\Delta \al_n$ suggest
that this ratio should not vary much as $\Delta \al_n$ is varied,
but, on the other hand the effect of $\Delta \be_n$ would be
magnified compared to what we show in Fig.~\ref{fig:dcsbeta}.

In summary, the extraction of ${\al_n}$ and ${\be_n}$ is possible
through coherent Compton scattering on unpolarized $\, ^3$He.
Extraction of ${\al_n}$ and ${\be_n}$ in this manner would also
serve as a check for the extractions of ${\al_n}$ and ${\be_n}$ from
the unpolarized $\ga$d experiments at MAXLab.

\subsection{Polarized Compton Scattering}

\begin{figure}[!htb]
\begin{center}
\includegraphics*[width=.48\linewidth]{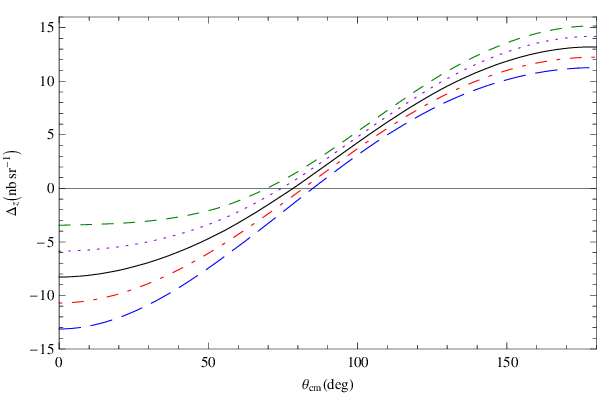}
\hfill
\includegraphics*[width=.48\linewidth]{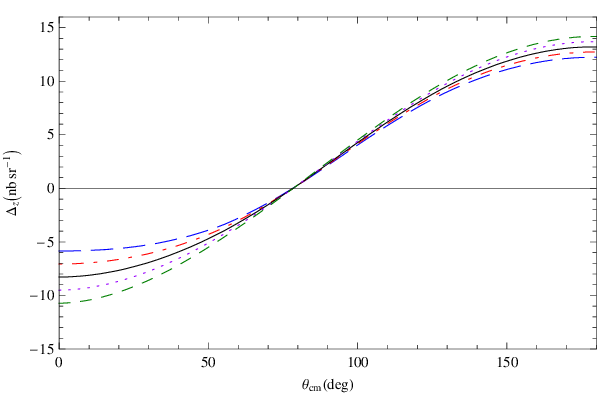}\\
\includegraphics*[width=.48\linewidth]{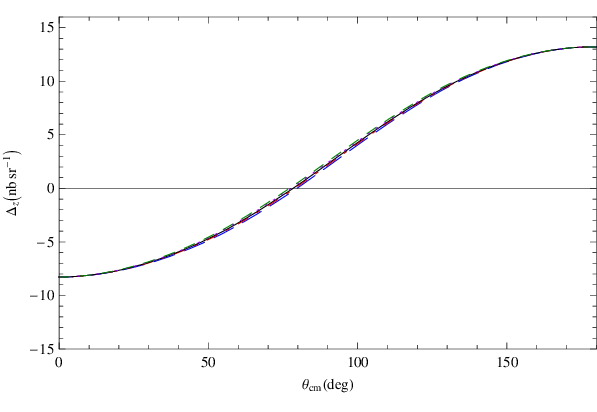}
\hfill
\includegraphics*[width=.48\linewidth]{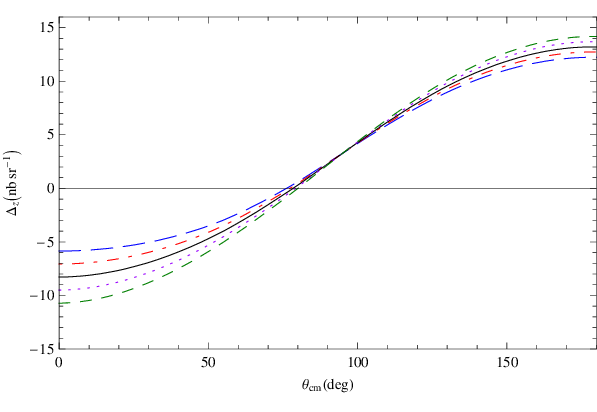}
\parbox{1.\textwidth}{
  \caption {The four panels above
correspond to $\Delta_z$ with varying each of $\Delta \gamma_{1n}$
(top left), $\Delta \gamma_{2n}$ (top right), $\Delta \gamma_{3n}$
(bottom left) and $\Delta \gamma_{4n}$ (bottom right), one at a
time. The calculations are done in c.m. frame at 120 MeV. The solid
(black) curves correspond to the full $\calO(e^2 Q)$ results. The
long-dashed (blue) curves correspond to $\Delta \ga_{i n} = - \ga_{i
n} (\calO(e^2 Q))$, dot-dashed (red) to $\Delta \ga_{i n} = - \ga_{i
n} (\calO(e^2 Q))/2$, dotted (magenta) to $\Delta \ga_{i n} = \ga_{i
n} (\calO(e^2 Q))/2$ and dashed (green) to $\Delta \ga_{i n} =
\ga_{i n} (\calO(e^2 Q))$.}
  \label{fig:ddcsz}}
\end{center}
\end{figure}

In Fig.~\ref{fig:ddcsz} we plot $\Delta_z$ vs. the c.m. angle at 120
MeV and the different panels correspond to varying the four spin
polarizabilities one by one. Here we choose to vary $\ga_{i n}, i=1
\ldots 4$ between $\pm 100\%$ of their $\calO(e^2 Q)$ prediction
(Eqs.~(\ref{eq:gamma4OQ3})). Looking at the plots, the message one
gets is that this observable is quite sensitive to $\ga_{1n}$,
$\ga_{2n}$ and $\ga_{4n}$. Especially with the expected increase in
photon flux at HI$\vec \ga$S such differences in cross-sections
should easily be measured and we can expect to extract a linear
combination of $\ga_{1n}$, $\ga_{2n}$ and $\ga_{4n}$. The linear
combination that readily comes to mind is $\ga_{0n}$ or $\ga_{\pi
n}$ but we should bear in mind that we are measuring $\Delta_z$ as a
function of angle which means that we should be able to extract the
combination $\ga_{1n} - (\ga_{2n} + 2\ga_{4n})\cos \theta$ (see
Eq.(\ref{eq:Asinw})). Observing the different plots in
Fig.~\ref{fig:ddcsz} we can already see the effect of the $\cos
\theta$ term in the panels for $\ga_{2n}$ and $\ga_{4n}$.
\begin{figure}[!htb]
\begin{center}
\includegraphics*[width=.48\linewidth]{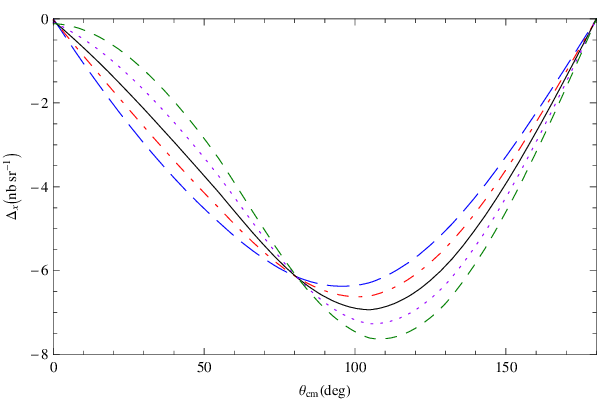}
\hfill
\includegraphics*[width=.48\linewidth]{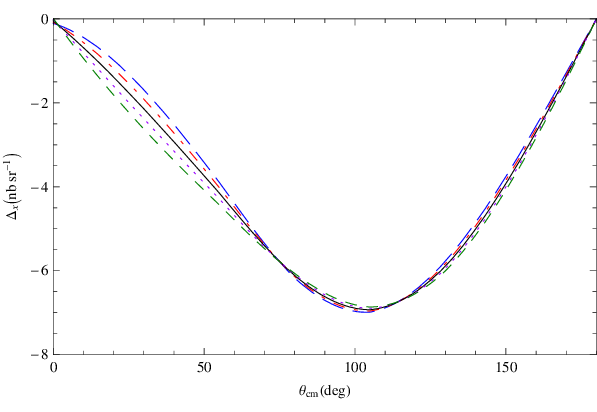}\\
\includegraphics*[width=.48\linewidth]{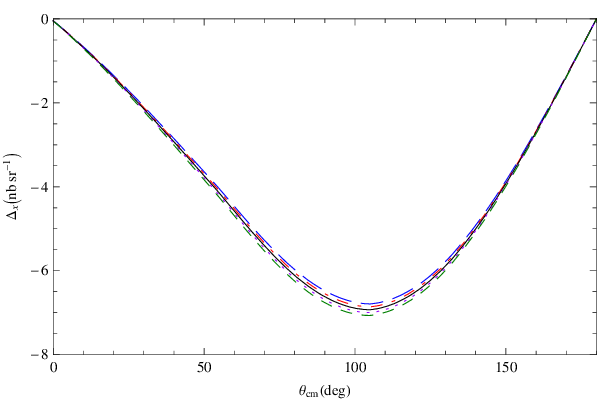}
\hfill
\includegraphics*[width=.48\linewidth]{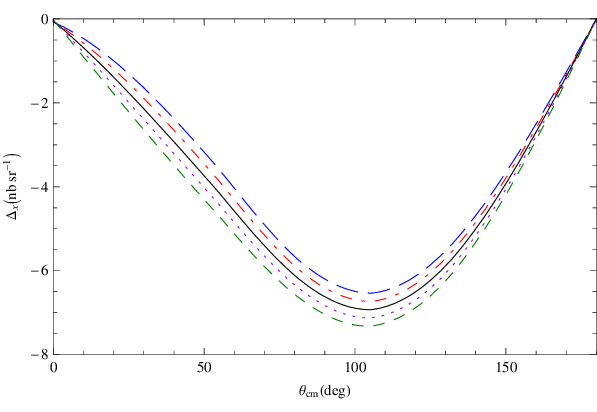}
\parbox{1.\textwidth}{
  \caption {he four panels above
correspond to $\Delta_x$ with varying each of $\Delta \gamma_{1n}$
(top left), $\Delta \gamma_{2n}$ (top right), $\Delta \gamma_{3n}$
(bottom left) and $\Delta \gamma_{4n}$ (bottom right), one at a
time. The calculations are done in c.m. frame at 120 MeV. The solid
(black) curves correspond to the full $\calO(e^2 Q)$ results. The
long-dashed (blue) curves correspond to $\Delta \ga_{i n} = - \ga_{i
n} (\calO(e^2 Q))$, dot-dashed (red) to $\Delta \ga_{i n} = - \ga_{i
n} (\calO(e^2 Q))/2$, dotted (magenta) to $\Delta \ga_{i n} = \ga_{i
n} (\calO(e^2 Q))/2$ and dashed (green) to $\Delta \ga_{i n} =
\ga_{i n} (\calO(e^2 Q))$.}
  \label{fig:ddcsx}}
\end{center}
\end{figure}

Next, in Fig.~\ref{fig:ddcsx} we plot $\Delta_x$ vs. the c.m. angle
at 120 MeV and the different panels correspond to varying the four
spin polarizabilities one by one. This figure also suggests that we
are sensitive to a combination of the same spin polarizabilities as
in $\Delta_z$ but this combination is obviously not the same as
before $i.e.$ $\ga_{1n} - (\ga_{2n} + 2\ga_{4n})\cos \theta$. We can
say that because the curves in the right panels of
Fig.~\ref{fig:ddcsx} should coincide at 90~deg because of the $\cos
\theta$ term but they do not. Hence, we should be able to extract a
different linear combination of the $\ga_n$s from $\Delta_x$.

Thus, we are sensitive to two different linear combinations of
$\ga_{1n}$, $\ga_{2n}$ and $\ga_{4n}$ through $\Delta_z$ and
$\Delta_x$ and at least we can expect that we can extract one, say
$\ga_{1n}$ unambiguously, out of the three and put constraints on
the values of the other two. For $\vec{\ga} \vec d$ scattering we
found that we were sensitive to a combination of $\ga_{1n}$ and
$\ga_{3n}$~\cite{Ch05}. This means that if we combine all the
information, we should be able to extract at least two of the four
spin polarizabilities and constrain the remaining two. Also we can
definitely hope that through polarized Compton scattering on
deuteron and \he3~ we can perform a nuclei independent extraction of
at least one of the neutron spin polarizabilities, presumably
$\ga_{1n}$.

It is worth mentioning here that the curves in Figs.~\ref{fig:ddcsz}
and \ref{fig:ddcsx} look very similar to those obtained if one
calculates the reaction $\vec \gamma \vec n \rightarrow \gamma n$,
i.e. Compton scattering off a free neutron. This would then suggest
that polarized \he3~ indeed behaves as an ``effective" neutron
target.

\section{Polarized \he3~ is Interesting}
\label{interesting}

In this section we would like to point out a couple of very
interesting facts about Compton scattering on polarized $\, ^3$He.
\begin{enumerate}

\item We know that \he3~ is a spin-${1\over 2}$ target and hence, we should be
able to write the photon scattering amplitude as a sum of six
structure functions like in Eq.(\ref{eq:Ti}). However in this case
these six functions have to be a sum of one-body and two-body parts
and are not as simple as the $A_i$s in Eq.~(\ref{eq:Ti}).
\begin{eqnarray}
T_{\ga ^3He}&=& \sum \limits_{i=1 \ldots 6} A_i^{^3He} t_i, \nonumber \\
A_i^{^3He} &=& A_i^{1B}+A_i^{2B}. \label{aihe3}
\end{eqnarray}
Here, $A_i^{1B}$ are the same as in Eq.(\ref{eq:Ti}) to the extent
that $^3$He is an ``effective" neutron. Meanwhile, we found that,
numerically, $A_i^{2B}, i=3 \ldots 6$ are negligible. Hence we
conclude that $A_i^{2B}$ does not contribute to the spin structure
functions at $\calO(e^2 Q)$. Moreover, since the two protons (to a
good approximation) have spins anti-aligned in a spin-polarized $\,
^3$He target, we can safely assume that all the sensitivity to the
spin polarizabilities in the observables come from the unpaired
neutron alone. This reasoning supports our claim (made in the
previous section) that the curves for \he3~ Compton scattering
resemble those obtained from $n$-Compton scattering. Thus, Compton
scattering on $\, ^3$He is an ideal avenue to extract the neutron
spin polarizabilities.

\item Let us consider the double polarization observables,
$\Delta_z$ and $\Delta_x$. If we take Eq.~(\ref{aihe3}) to calculate
these observables then we shall obtain expressions similar to
Eqs.~(3.19) and (3.22) of Ref.~\cite{mythesis} but with $A_i$
replaced by $A_i^{^3He}$. From these equations we know that to the
lowest order in $\w$ (${\mathcal O}(\w^3)$) the spin
polarizabilities manifest themselves via the interference terms of
$A_3 \ldots A_6$ with $A_1$. Hence, to the extent that polarized $\,
^3$He is an effective neutron, we can expect the $A_3 \ldots A_6$
for the neutron will be multiplied by $A_1^{^3He}$, which-- at least
at low energies-- is dominated by the two proton Thomson terms, thus
giving a much enhanced effect compared to $\ga$d.

\end{enumerate}

\section{Sources of Uncertainties}
\label{sec:err}

\subsection{Wavefunction Dependence}

\begin{figure}[!htb]
\begin{center}
\includegraphics*[width=.48\linewidth]{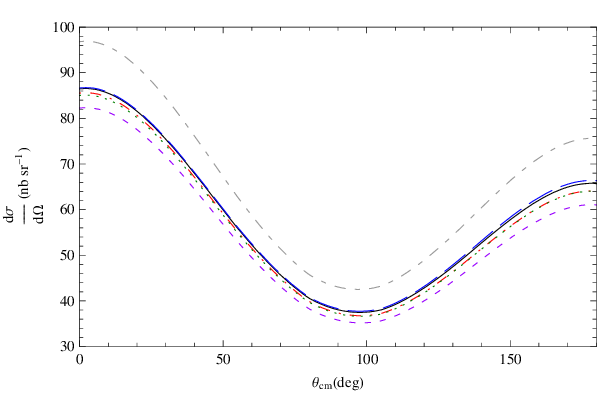}
\hfill
\includegraphics*[width=.48\linewidth]{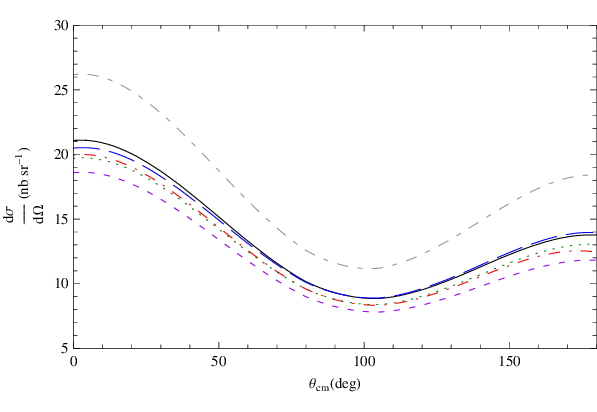}\\
\includegraphics*[width=.48\linewidth]{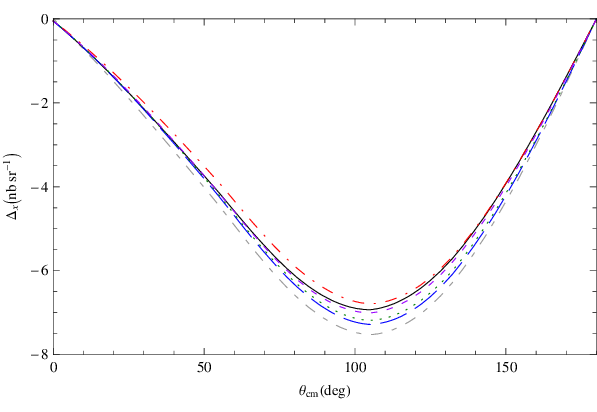}
\hfill
\includegraphics*[width=.48\linewidth]{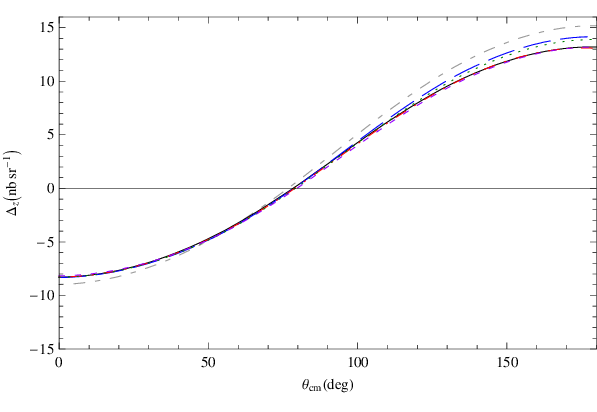}
\parbox{1.\textwidth}{
  \caption {The top two panel show the dcs at 60 MeV (left) and 120 MeV (right) calculated with different wavefunctions.
  The bottom left panel corresponds to $\Delta_x$ at 120 MeV and the bottom right panel corresponds to $\Delta_z$ at 120 MeV.
   These have also been calculated using different wavefunctions. The dashed-shortdashed (grey) curves correspond to $AV18 + UrbIX$, solid (black) corresponds to
   Idaho $N^3LO + 3NFB$, the long-dashed (blue) curves correspond to Idaho $N^3LO + 3NFA$, the dot-dashed (red)
   curves correspond to  Idaho $N^3LO$ with $\Lambda = 500$, the dotted
   (green) curves
   correspond to  Idaho $N^3LO$ with $\Lambda = 600$ and the dashed (magenta) curves correspond to
   $NLO$ chiral wavefunction with $\Lambda=600$. All the calculations have been done in the c.m. frame.}
  \label{fig:wfdep}}
\end{center}
\end{figure}

In calculating the matrix elements (Eq.~(\ref{eq1})), ideally, the
wavefunctions and the operator should be calculated within the same
chiral framework, so that the formalism is consistent. Although
few-nucleon bound state wavefunctions based on chiral nuclear
interactions have been calculated, the currently used approaches are
not entirely consistent with the approach used for the Compton
scattering operators. Therefore, we employ several wavefunctions
based on chiral perturbation theory and additional wavefunction
obtained from the state-of-the-art interaction models. In this way
we expect to cover the entire dependence on the choice for the
wavefunctions. Specifically we chose-
\begin{enumerate}
\item The phenomenological potential $AV18$~\cite{Wi95} with the $3N$
interaction Urbana IX~\cite{Pu95}.

\item The Idaho $N^3LO$ chiral potential~\cite{En03} with the cut-off $\Lambda =
500$~MeV with a chiral $3N$ force~\cite{No06} (called $3NFB$ here).

\item The Idaho $N^3LO$ chiral potential~\cite{En03} with the cut-off $\Lambda =
500$~MeV with another chiral $3N$ force~\cite{No06} (called $3NFA$
here). The difference between $3NFA$ and $3NFB$ is that they are
parameterized differently so that they reproduce the triton and
alpha-particle binding energy but lead to different spectra for
heavier nuclei~\cite{No06}.

\item The Idaho $N^3LO$ chiral potential~\cite{En03} with the cut-off $\Lambda =
500$~MeV and no $3NF$.

\item The Idaho $N^3LO$ chiral potential~\cite{En03} with the cut-off $\Lambda =
600$~MeV and no $3NF$.

\item The $NLO$ chiral potential with dimensional regularization~\cite{Ep02} with the cut-off in the Lippman-Schwinger set to $\Lambda =
600$~MeV and no $3NF$.

\end{enumerate}
In the future, it will also be interesting to employ wavefunctions
based on $NLO$, $N^2LO$ and $N^3LO$ interactions of~\cite{Ep05},
which use a different regularization scheme and allow one to compare
several orders of the chiral interaction within the one framework.
However, we believe that the set of interaction from above should
give a reasonable idea of possible wavefunction dependence in this
exploratory study. What we want to demonstrate through the choice of
these particular wavefunctions is that-
\begin{itemize}
\item A contrast between the $NLO$ chiral wavefunction ($\Lambda=600$~MeV)
and the Idaho $N^3LO$ wavefunction ($\Lambda=600$~MeV) will
demonstrate the effect of terms of higher chiral order.

\item A contrast between the Idaho $N^3LO$ wavefunction
($\Lambda=600$~MeV) without a $3NF$ and the Idaho $N^3LO$
wavefunction ($\Lambda=500$~MeV) will demonstrate the effect of
varying the cut-off which translates into studying the impact of
short-distance physics.

\item A contrast between the Idaho $N^3LO$ wavefunction ($\Lambda=500$~MeV)
and the Idaho $N^3LO$ wavefunction ($\Lambda=500$~MeV) with $3NFA$
will demonstrate the effect of `a' three-nucleon force.

\item A contrast between the Idaho $N^3LO$ wavefunction
($\Lambda=500$~MeV) with $3NFA$ and the Idaho $N^3LO$ wavefunction
($\Lambda=500$~MeV) with $3NFB$ will demonstrate the effect of
choosing three-nucleon forces with different parameters.

\item A contrast between the Idaho $N^3LO$ wavefunction
($\Lambda=500$~MeV) with $3NFB$ and the $AV18+$Urbana-IX
wavefunction will demonstrate the effect of choosing a wavefunction
constructed with a phenomenological potential.

\end{itemize}
We have calculated the dcs at two different energies (60 \& 120~MeV)
and also $\Delta_x$ and $\Delta_z$ at 120~MeV with these six
wavefunctions. The results are plotted in Fig.~\ref{fig:wfdep}.

Let us first compare the top two panels that show the dcs. The first
thing to notice is that the shape of the dcs is the same no matter
what wavefunction we use. Another fact that stands out is that all
the chiral potentials produce dcses that are close to one another,
whereas the choice of $AV18+$Urbana-IX causes the dcs to be quite
different. Also the size of the difference in the dcs due to the
choice of different wavefunctions increases as we go from 60 to
120~MeV. One major difference between the chiral wavefunctions and
the $AV18+$Urbana-IX wavefunction is that the latter has explicit
high-momentum components in the NN interaction and this causes the
dcs also to be significantly different. Having said that, it is
essential to acknowledge the fact that the dcs is sensitive to the
choice of different chiral potentials too. Of these, the chiral
wavefunctions with three-nucleon forces and the $N^3LO$ chiral
wavefunctions are actually of a higher chiral order than our Compton
scattering operator and we expect some sensitivity to the higher
chiral order terms. Nevertheless, looking at the dcs it can be said
that the uncertainty due to the choice of wavefunctions (chiral
wavefunctions only) is $\lesssim 15\%$ at 120~MeV. Of course this is
lower at 60~MeV ($\lesssim 8\%$) and we advocate that if a
measurement of the dcs is necessary to extract $\al_n$ or $\be_n$ it
should be done at lower energies.

Next, observe the lower panels of Fig.~\ref{fig:wfdep}. What
immediately stands out is that even though these calculations were
done at 120~MeV, the sensitivity to the choice of wavefunctions is
much smaller than in the dcs. This is a good sign because we want to
use these double-polarization observables to extract the $\ga_n$s.
The uncertainty due to the choice of different wavefunctions in
these observables is $\lesssim 7.5\%$. Also, the calculations with
the different chiral potentials converge to a single value ($i.e.$
uncertainty is $\sim 0\%$) at forward angles for $\Delta_z$ and this
augurs well for us because (as we see in Fig.~\ref{fig:ddcsz}) there
is maximum sensitivity to the $\ga_n$s at forward angles. Similarly,
a measurement of $\Delta_x$ at forward angles will also reduce the
effect of the wavefunction dependence. But, it should be pointed out
that, experimentally, at forward angles there are large background
effects and it will be difficult to separate the $\ga \, ^3$He
scattering events from the background at HI$\vec{\ga}$S.

\subsection{Position of Pion-production Threshold}

In HB\cpt it is assumed that the nucleon rest four momenta dominate
their total four momenta, in other words, the nucleons are virtually
at rest. In consequence the pion-production threshold occurs at
$\w=\mpi$, because the nucleus cannot recoil. Hence, this assumption
creates a problem for all processes involving baryons near the
pion-production threshold. The position of the pion-production
threshold in $\ga \, ^3$He scattering is actually at a photon energy
of
\begin{equation}
\w = -B_{^3He} + \mpi + \frac{\mpi^2}{2M_{^3He}} + {\mathcal
O}(\mpi^3),
\end{equation}
and not at $\w=\mpi$. Hence, for the purpose of understanding the
effect of this in our calculations, we redefined
\begin{equation}
\tilde{\w} = -B_{^3He} + \w + \frac{\w^2}{2M_{^3He}}
\label{eq:wtildehe3}
\end{equation}
as the energy `going into' the $\ga N$ amplitude and calculated the
dcs at $\w=$100~MeV. $B_{^3He}$ is the binding energy of the $^3$He
nucleus in Eq.~(\ref{eq:wtildehe3}). We found that there was a $\sim
5\%$ difference in the dcs at 100~MeV. In $\Delta_x$ the difference
was $\sim 1\%$ and in $\Delta_z$ the difference was $\sim 3\%$ at
100~MeV. It is encouraging that the percentage uncertainty in the
double polarization observables is again less than that in the dcs.
This investigation was just exploratory and a proper treatment of
this issue is an important topic for future study.

\subsection{Boost Corrections}

Our calculation of the $\ga$\he3~ scattering process is in the $\ga
{}^3$He c.m. frame. However, the one-body amplitude
(Eq.~(\ref{eq:Ti})) is defined to be in the $\ga $N c.m. frame and
similarly the two-body amplitude is defined to be in the $\ga$NN
c.m. frame. Thus, we need to boost these amplitudes to the $\ga$NNN
c.m. frame. This is necessary to ensure that the vectors $\veps$ and
$\vepsprime$ stay orthogonal to $\vk$ and $\vkp$ respectively in the
new frame. Consequently, a `boosting' term has to be added to the
$\ga $N amplitude which is given by-
\begin{eqnarray}
{\hat O}_{Boost}^{(1B)}&=& \frac{e^2}{M^2 \w} \left[ 1+ {1\over 2}
(\tau_3^{(1)}+\tau_3^{(2)}) \right] \nonumber \\
&&\left[ {1\over 3} \vec {\epsilon}'\cdot \vec k  \,  \vec
{\epsilon} \cdot \vec k' + {1\over 2} (\vec {\epsilon}'\cdot \vec
{p_3}  \, \vec {\epsilon} \cdot \vec k' + \vec {\epsilon} \cdot \vec
{p_3}  \, \vec
{\epsilon}' \cdot \vec k) \right] \nonumber \\
&-& {1\over 2}\frac{e^2}{M^2 \w} (\tau_3^{(1)}-\tau_3^{(2)})[\vec
{\epsilon}'\cdot \vec {p_{12}} \,  \vec {\epsilon} \cdot \vec k' +
\vec {\epsilon} \cdot \vec {p_{12}} \, \vec {\epsilon}' \cdot \vec k
]. \label{boost}
\end{eqnarray}
Here, all symbols have their usual meanings. The structure of the
two-body operator is such that it is Galilean boost invariant. We
calculated the effect of this boost at 100~MeV and found that it is
truly a perturbative effect and the difference is $\lesssim 0.01\%$
in the dcs. This is because the boost corrections to the transition
matrix elements that dominate are really very small if not zero. The
boost pieces contribute to those matrix elements that are themselves
small and hence play only a small role in the observables.

\section{Summary and Future Prospects}
\label{sec:sum}

We have performed the first ${\mathcal O}(e^2 Q)$ HB$\chi$PT
calculations of Compton scattering on \he3. Indeed we are not aware
of any previous theoretical calculation of this reaction process in
\underline{any} framework. We have shown that \he3~ Compton
scattering is an exciting prospect for mining the neutron
polarizabilities. Polarized \he3~ is unique and interesting because
it does seem to behave as an ``effective neutron" (refer to
Sec.~\ref{interesting}) in Compton scattering. With the added
advantage of a significant Compton cross-section due to the presence
of the two protons, whose Thomson terms can interfere with the
neutron polarizabilities, we get a significantly magnified effect
compared to $\ga$d. The results reported in Sec.~\ref{sec:res} are
quite encouraging because they show that not only can we access
$\al_n$ and $\be_n$ through the dcs but we can also hope to extract
two different linear combinations of $\ga_{1n}$, $\ga_{2n}$ and
$\ga_{4n}$ from the double-polarization observables, $\Delta_z$ and
$\Delta_x$.

Meanwhile, priorities for HI$\vec \ga$S (at Triangle Universities
Nuclear laboratory) were discussed at a Compton workshop held prior
to the Fifth International Workshop on Chiral Dynamics,
2006~\cite{Ha06} and they included polarized $\vec{\ga} \vec p$,
$\vec{\ga} \vec d$ and $\vec{\ga} \vec{ ^3He}$ experiments. The idea
is that the $\vec \ga \vec p \rightarrow \ga p$ experiments will
reveal $\ga_{1p} \ldots \ga_{4p}$ and the set for the proton
polarizabilities will be complete. This will set the stage for the
neutron polarizabilities. Moreover, if the MAXLab experiments
generate enough data to extract $\al_n$ and $\be_n$ unambiguously,
then $\ga {^3}$He can be used to focus on the neutron spin
polarizabilities. One can also hope that coherent ~\he3 Compton
scattering measurements may be done at MAXLab in the future. The
expectation is that, at the completion of all these planned
experiments we will be able to complete the set of polarizabilities
for the neutron.

However, there is a lot of scope for improving the theoretical
calculations on \he3~ Compton scattering. This being the first
calculation was exploratory in nature. Improvements on the
theoretical side could include higher-order (${\mathcal O} (e^2
Q^2)$) calculations and/or explicit inclusion of the
$\Delta-$isobar. (We have seen that explicit inclusion of the
$\Delta$-isobar causes noticeable changes in the results for
$\vec{\ga} \vec d$ scattering.) Also dedicated effort is required to
shrink the theoretical uncertainties associated with the choice of
nuclear wavefunctions. Hildebrandt $et\, al.$~\cite{Hi05b, Hi05}
have demonstrated that by resumming the intermediate NN scattering
states they were able to restore the Thomson limit for deuteron
Compton scattering and also significantly reduce the dependence on
the choice of deuteron wavefunctions. Such an exercise can also be
done for \he3~ Compton scattering. Effort is also required to
rigorously study the effect of the incorrect position of the
pion-production threshold due to the ``static approximation" for
nucleons in HB\cpt. These effects have been studied for $\pi d$
scattering by Baru $et \, al.$~\cite{Ba04} and for $\ga d
\rightarrow \pi^+ nn$ by Lensky $et \, al.$~\cite{Le05}. Using
similar principles, we can endeavor to resolve this issue for
Compton scattering on nuclear targets.

To conclude, it should be reiterated that we have taken the crucial
first step. Of course, much needs to be done but we are confident
that with sustained efforts we shall be able to extract the neutron
polarizabilities from forthcoming $\ga$\he3~ scattering data.

\section*{Acknowledgement}
We thank Haiyan Gao and Al Nathan for useful discussions. DS
expresses her gratitude to the Indian Institute of Technology at
Kanpur (India) where a part of this work was completed. This work
was carried out under grants DE-FG02-93ER40756 of US-DOE (DS and
DRP) and PHY-0645498 (DS) of NSF. The wave functions have been
computed at the NIC, J\"ulich.

%%%%%%%%%%%%%%%%%%%%%%%%%%%%%%%%%%%%%%%%%%%%%%%%%%%%%%%%%%%
\newpage
\appendix

\section{$\ga$N Amplitude}
\label{sec:gaN}
\setcounter{equation}{0}

After evaluating all the diagrams in Figs.~\ref{fig:tree} and
\ref{fig:IA}, it becomes evident that the photon scattering
amplitude $T_{\gamma N}$ is a sum of six operator structures that
can be expressed as:
\begin{eqnarray}
T_{\gamma N}&=& e^2\{ A_1 \vepsprime\cdot\veps
             +A_2 \vepsprime\cdot\vk \, \veps\cdot\vkp
             +iA_3 \vsigma\cdot (\vepsprime\times\veps) \nonumber \\
            & & + iA_4 \vsigma\cdot (\vkp\times\vk)\, \vepsprime\cdot\veps
             +iA_5 \vsigma\cdot [(\vepsprime\times\vk)\, \veps\cdot\vkp
            -(\veps\times\vkp)\, \vepsprime\cdot\vk] \nonumber \\
           & &+ iA_6 \vsigma\cdot [(\vepsprime\times\vkp)\, \veps\cdot\vkp
                        -(\veps\times\vk)\, \vepsprime\cdot\vk]
                        \} \\
           &=& \sum_{i=1}^6 A_i t_i
\label{eq:Ti}
\end{eqnarray}
where,
\begin{eqnarray}
t_1 &=& e^2 \vepsprime\cdot\veps \nonumber \\
t_1 &=& e^2 \vepsprime\cdot\vk \, \veps\cdot\vkp \nonumber \\
t_1 &=& i e^2 \vsigma\cdot (\vepsprime\times\veps) \nonumber \\
t_1 &=& i e^2 \vsigma\cdot (\vkp\times\vk)\, \vepsprime\cdot\veps \nonumber \\
t_1 &=& i e^2 \vsigma\cdot [(\vepsprime\times\vk)\, \veps\cdot\vkp
            -(\veps\times\vkp)\, \vepsprime\cdot\vk] \nonumber \\
t_6 &=& i e^2 \vsigma\cdot [(\vepsprime\times\vkp)\, \veps\cdot\vkp
                        -(\veps\times\vk)\, \vepsprime\cdot\vk] \nonumber \\
\end{eqnarray}
The six invariant amplitudes, $A_1 \ldots A_6$, are functions of the
photon energy, $\w$, and the Mandelstam variable $t$. Defining
$\barf =\omega/m_\pi$ and $t=-2{\barf ^2}(1-\cos \theta)$, where
$\theta$ is the center-of-mass angle between the incoming and
outgoing photon momenta, one finds \cite{bkmrev, Be91, Be92, Be93,
Be94}:
\begin{eqnarray}
A_1 &=& -\frac{{\mathcal Z}^2}{M}
        +\frac{g_A^2m_\pi}{8\pi f_\pi^2}
         \left\{ 1- \sqrt{1-\barf^2}
                 +\frac{2-t}{\sqrt{-t}}
                  \left[\frac{1}{2} \arctan \frac{\sqrt{-t}}{2}
                        -I_1(\barf,t) \right]\right\},
\nonumber \\
A_2 &=& \frac{{\mathcal Z}^2 \w}{M^2}
        -\frac{g_A^2 \w^2}{8\pi f_\pi^2 m_\pi}
         \frac{2-t}{(-t)^{3/2}}
         \left[I_1(\barf,t)- I_2(\barf,t)\right],
\nonumber \\
A_3 &=& \frac{\omega}{2M^2} [{\mathcal Z}({\mathcal Z}+2\kappa
)-({\mathcal Z}+\kappa)^2 \cos \theta]
        +\frac{{(2{\mathcal Z} -1)}g_Am_\pi}{8\pi^2 f_\pi^2} \frac{\barf t}{1-t}
\nonumber \\
    & &  +\frac{g_A^2m_\pi}{8\pi^2 f_\pi^2}
         \left[ \frac{1}{\barf} \arcsin^2\barf- \barf +2\barf^4
\sin^2 \theta I_3(\barf,t)\right],
\nonumber \\
A_4 &=& -\frac{({\mathcal Z}+\kappa )^2\omega}{2M^2}
        +\frac{g_A^2 \w^2}{4\pi^2 f_\pi^2m_\pi} I_4(\barf,t),
\nonumber \\
A_5 &=& \frac{({\mathcal Z}+\kappa )^2\omega}{2M^2}
        -\frac{{(2{\mathcal Z} -1)}g_A \w^2}{8\pi^2 f_\pi^2 m_\pi} \frac{\barf}
        {(1-t)}
        -\frac{g_A^2 \w^2}{8\pi^2 f_\pi^2m_\pi}
          [I_5(\barf,t)-2\barf^2\cos\theta I_3(\barf,t)],
\nonumber \\
A_6 &=& -\frac{{\mathcal Z}({\mathcal Z}+\kappa )\omega}{2M^2}
        +\frac{{(2{\mathcal Z} -1)}g_A \w^2}{8\pi^2 f_\pi^2 m_\pi}\frac{\barf}
         {(1-t)} \nonumber \\
        &&+\frac{g_A^2 \w^2}{8\pi^2 f_\pi^2m_\pi}
          [I_5(\barf,t)-2\barf^2 I_3(\barf,t)],
\label{eq:As}
\end{eqnarray}
where ${\mathcal Z}=1$(0) for the proton(neutron) and:
\begin{eqnarray}
I_1(\barf,t) &=& \int_0^1  dz \,
             \arctan \frac{(1-z)\sqrt{-t}}{2\sqrt{1-\barf^2 z^2}},
\nonumber \\
I_2(\barf,t) &=& \int_0^1  dz \,
             \frac{2(1-z)\sqrt{-t(1-\barf^2z^2)}}{4(1-\barf^2 z^2)-t(1-z)^2},
\nonumber \\
I_3(\barf,t) &=& \int_0^1  dx \, \int_0^1  dz \,
             \frac{x(1-x)z(1-z)^3}{S^3}
             \left[ \arcsin \frac{\barf z}{R}+ \frac{\barf zS}{R^2}\right],
\nonumber \\
I_4(\barf,t) &=& \int_0^1  dx \, \int_0^1  dz \,
             \frac{z(1-z)}{S}\arcsin \frac{\barf z}{R},
\nonumber \\
I_5(\barf,t) &=& \int_0^1  dx \, \int_0^1  dz \,
             \frac{(1-z)^2}{S}\arcsin \frac{\barf z}{R},
\label{eq:Is}
\end{eqnarray}
with
\begin{equation}
S=\sqrt{1-\barf^2 z^2-t(1-z)^2x(1-x)}, \qquad
R=\sqrt{1-t(1-z)^2x(1-x)}. \label{eq:sr}
\end{equation}

\section{$\ga$NN Amplitude}
\label{sec:gaNN}
\setcounter{equation}{0}

The two-body diagrams that contribute to the Compton scattering
process at ${\mathcal O}(e^2 Q)$ are shown in Fig.~\ref{fig:2B}. The
two-body amplitude can be expressed (in the $\gamma$NN c.m.
frame)~\cite{Be99, Be02, Be04} as:
\begin{equation}
T_{\gamma NN}^{2B}\;=\;-\frac{{e^2}{g_A^2}}{2 f_\pi^2} \;
({\vec\tau}^{\; 1} \cdot{\vec\tau}^{\; 2}-\tau^{1}_{3}\tau^{2}_{3})
\; ( t^{(a)}+ t^{(b)}+ t^{(c)}+ t^{(d)}+ t^{(e)}), \label{eq:total}
\end{equation}
where
\begin{eqnarray}
{t^{(a)}}&= & \frac{{\veps\cdot\vsigone}\;{\vepsprime\cdot\vsigtwo}}
{2\lbrack {\omega^2}-{m_\pi^2}- (\vpee -\vpeeprime
+{\frac{1}{2}(\vkay +\vkayprime )})^2 \rbrack} +
(1\;\leftrightarrow\; 2),
\label{ta}\\
{t^{(b)}}&= & \frac{{\veps\cdot\vepsprime}\; \vsigone\cdot ( \vpee
-\vpeeprime -{\frac{1}{2}(\vkay -\vkayprime )} )
        \vsigtwo\cdot ( \vpee -\vpeeprime +{\frac{1}{2}(\vkay -\vkayprime )})}
{2\lbrack (\vpee -\vpeeprime -{\frac{1}{2}(\vkay -\vkayprime )})^2
+{m_\pi^2} \rbrack \lbrack (\vpee -\vpeeprime +{\frac{1}{2}(\vkay
-\vkayprime )})^2 +{m_\pi^2} \rbrack} \nonumber \\
&+& (1\;\leftrightarrow\; 2),
\label{tb}\\
{t^{(c)}}&= & -\frac{{\vepsprime\cdot (\vpee -\vpeeprime
+{\frac{1}{2}\vkay})}\;
          \vsigone\cdot\veps\;
 \vsigtwo\cdot ( \vpee -\vpeeprime +{\frac{1}{2}(\vkay -\vkayprime )} )}
{\lbrack {\omega^2}-{m_\pi^2}- (\vpee -\vpeeprime
+{\frac{1}{2}(\vkay +\vkayprime )})^2 \rbrack \lbrack (\vpee
-\vpeeprime +{\frac{1}{2}(\vkay -\vkayprime )})^2
+{m_\pi^2} \rbrack} \nonumber \\
&+& (1\;\leftrightarrow\; 2),
\label{tc}\\
{t^{(d)}}&= & -\frac{{\veps\cdot (\vpee -\vpeeprime
+{\frac{1}{2}\vkayprime} )}\; \vsigone\cdot ( \vpee -\vpeeprime
-{\frac{1}{2}(\vkay -\vkayprime )} )
              \vsigtwo\cdot\vepsprime }
{\lbrack {\omega^2}-{m_\pi^2}- (\vpee -\vpeeprime +
{\frac{1}{2}(\vkay +\vkayprime )})^2 \rbrack \lbrack (\vpee
-\vpeeprime -{\frac{1}{2}(\vkay -\vkayprime )})^2
+{m_\pi^2} \rbrack} \nonumber \\
&+& (1\;\leftrightarrow\; 2),
\label{td}\\
{t^{(e)}}&= & \frac{2 {\veps\cdot (\vpee -\vpeeprime
+{\frac{1}{2}\vkayprime})\; \vepsprime\cdot (\vpee -\vpeeprime
+{\frac{1}{2}\vkay})}\; \vsigone\cdot ( \vpee -\vpeeprime
-{\frac{1}{2}(\vkay -\vkayprime )})\; } {\lbrack
{\omega^2}-{m_\pi^2}- (\vpee -\vpeeprime + {\frac{1}{2}(\vkay
+\vkayprime )})^2 \rbrack \lbrack (\vpee -\vpeeprime
-{\frac{1}{2}(\vkay -\vkayprime )})^2
+{m_\pi^2} \rbrack}\nonumber \\
&\times& \frac{\vsigtwo\cdot ( \vpee -\vpeeprime +{\frac{1}{2}(\vkay
- \vkayprime )} )}{\lbrack (\vpee -\vpeeprime +{\frac{1}{2}(\vkay
-\vkayprime )}) ^2  +{m_\pi^2} \rbrack} + (1\;\leftrightarrow\; 2),
\label{te}
\end{eqnarray}
Here again, $\veps, \vepsprime, \vkay, \vkayprime$ have their usual
meaning. $\vpee$($\vpeeprime$) is the initial (final) momentum of
the nucleon inside the deuteron, and $\vsigone$($\vsigtwo$) is twice
the spin operator of the first (second) nucleon. The numbering of
the nucleons is arbitrary and hence ($1\leftrightarrow 2$) denotes
the term when the nucleons are interchanged.

%%%%%%%%%%%%%%%%%%%%%%%%%%%%%%%%%%%%%%%%%%%%%%%%%%%%%%%%%%%%%%%%%%%%%%%%

\end{document}